\documentclass[twocolumn]{aastex631}

\usepackage{natbib}
\usepackage{aas_macros}
\usepackage{hyperref}
\usepackage{academicons}
\usepackage{xcolor}

\usepackage{graphicx}

\begin{document}

\title{The Halo Occupation Distribution Modeling of the X-ray-selected AGNs at $0.6<\MakeLowercase{z}<2.6$ \\ in the COSMOS field}
\author[0000-0002-1207-1979]{Hiroyuki Ikeda}
\affiliation{National Institute of Technology, Wakayama College, Gobo, Wakayama 644-0023, Japan}
\author[0000-0002-7562-485X]{Takamitsu Miyaji}
\affiliation{Instituto de Astronom\'ia, Universidad Nacional Aut\'onoma de M\'exico, AP 106, Ensenada, Baja California, 22800 Mexico}
\author{Taira Oogi}
\affiliation{Department of Electrical and Computer Engineering, National Institute of Technology, Asahikawa College, Shunkodai 2-2-1-6, Asahikawa, Hokkaido 071-8142, Japan}
\author[0000-0002-3531-7863]{Yoshiki Toba}
\altaffiliation{NAOJ fellow}
\affiliation{National Astronomical Observatory of Japan, 2-21-1 Osawa, Mitaka, Tokyo 181-8588, Japan}
\affiliation{Academia Sinica Institute of Astronomy and Astrophysics, 11F of Astronomy-Mathematics Building, AS/NTU, No.1, Section 4, Roosevelt Road, Taipei 10617, Taiwan}
\affiliation{Research Center for Space and Cosmic Evolution, Ehime University, 2-5 Bunkyo-cho, Matsuyama, Ehime 790-8577, Japan}
\author[0000-0002-7348-8815]{H\'{e}ctor Aceves }
\affiliation{Instituto de Astronom\'ia, Universidad Nacional Aut\'onoma de M\'exico, AP 106, Ensenada, Baja California, 22800 Mexico}
\author[0000-0001-5544-0749]{Stefano Marchesi}
\affiliation{Dipartimento di Fisica e Astronomia (DIFA), Università di Bologna, via Gobetti 93/2, I-40129 Bologna, Italy}
\affiliation{Department of Physics and Astronomy, Clemson University, Kinard Lab of Physics, Clemson, SC 29634-0978, USA}
\affiliation{INAF - Osservatorio di Astrofisica e Scienza dello Spazio di Bologna, Via Piero Gobetti, 93/3, 40129, Bologna, Italy}
\author[0000-0001-7232-5152]{Viola Allevato}
\affiliation{INAF - Osservatorio Astronomico di Capodimonte Salita Moiariello 16, 80131, Napoli, Italy}
\author[0000-0001-9383-786X]{Akke Viitanen}
\affiliation{INAF–Osservatorio Astronomico di Roma, via Frascati 33, 00040 Monteporzio Catone, Italy}
\affiliation{Department of Physics, University of Helsinki, PO Box 64, FI-00014 Helsinki, Finland}
\author[0000-0002-2115-1137]{Francesca Civano}
\affiliation{Astrophysics Science Division, NASA Goddard Space Flight Center, Greenbelt, MD 20771, USA}

\email{h-ikeda@wakayama-nct.ac.jp}
\email{ikedahr0@gmail.com}


\begin{abstract}
We conducted precise measurements of Active Galactic Nuclei (AGNs) clustering at $z\sim1$ and $z\sim2$ by measuring the two-point cross-correlation function (CCF) between galaxies and X-ray-selected AGNs, and the two-point auto-correlation function (ACF) of galaxies in the COSMOS field to interpret the CCF results. The galaxy sample was selected from the COSMOS2015 catalog, while the AGN sample was chosen from the {\sl Chandra} COSMOS-Legacy survey catalog.
For the AGN samples at $z\sim1$ and $z\sim2$, we calculated AGN bias values of $b=1.16\ (1.16;1.31)$ and $b=2.95\ (2.30;3.55)$, respectively. These values correspond to typical host dark matter halo (DMH) masses of log$(M_{\rm typ}/M_{\odot})=11.82\ (11.82;12.12)$ and log$(M_{\rm typ}/M_{\odot})=12.80\ (12.38;13.06)$, respectively. 
Subsequently, we performed Halo Occupation Distribution (HOD) modeling of X-ray-selected AGNs using the CCF and ACF of galaxies. 
We have found a significant satellite AGN population at $z\sim 1$ all over the DMH mass ($M_{\rm DMH}$) range occupied by AGNs. While $z\sim 2$ AGNs in our sample are associated with higher mass DMHs and smaller satellite fractions.
The HOD analysis suggests a marginal tendency of increasing satellite slope with redshift, but larger samples are needed to confirm this with sufficient statistical significance. We find that the best-fit values of satellite slope in both redshift bins are greater than 0, suggesting tendencies of increasing satellite AGN number with $M_{\rm DMH}$. 

 \end{abstract}


\section{Introduction}
Previous observations have shown that a small fraction of galaxies exhibit active galactic nuclei (AGNs) activity. It is well established that almost all galaxies have a central supermassive black hole (SMBH), implying that the galaxies had gone through at least one period of AGN activity, i.e., accretion of material onto the nucleus, in the past. The conditions and mechanisms in which AGN activity occurs have been the subject of intensive research. Observational clues to tackle this problem include the luminosity function of AGNs and its evolution (\citealt{2003ApJ...598..886U,2014ApJ...786..104U,2005A&A...441..417H,2009MNRAS.399.1755C,2011ApJ...728L..25I,2015ApJ...804..104M,2015MNRAS.451.1892A,2016A&A...587A.142F,2016A&A...590A..80R,2012ApJ...756..160I,2015ApJ...798...28K,2016ApJ...829...33Y,2018PASJ...70S..34A,2018ApJ...869..150M,2019ApJ...871..240A,2020ApJ...904...89N,2023ApJ...949L..42M}), black hole (BH) demography and AGN host galaxy properties. AGN clustering gives alternative clues.
The clustering measurements of AGNs provide us with where AGNs are in the Universe. Therefore, the two-point auto-correlation function (ACF) of AGNs has been studied by many investigators \citep[e.g.,][]{2007ApJS..172..396M, 2008ApJS..179..124U,2009A&A...494...33G,2011ApJ...741...15S,2013MNRAS.428.1382K,2014ApJ...796....4A,2017ApJ...835...36T,2019A&A...632A..88A,2019A&A...629A..14V,krumpe23} as the most common measure of clustering. They have calculated the large-scale bias and found that the typical host dark matter halo (DMH) mass ($M_{\rm typ}$) is $\sim 10^{12.5}-10^{13.0}M_{\odot}$. 

When the number of AGNs with known spectroscopic redshift ($z_{\rm sp}$) is small, the two-point cross-correlation function (CCF) between galaxies and AGNs is often used  \citep[e.g.,][]{2005ApJ...627L...1A,2008ApJ...673L..13F,2009ApJ...701.1484C,2009MNRAS.394.2050M,2013ApJ...778...98S,2015ApJ...809..138I,2017ApJ...848....7G,2018PASJ...70S..33H,2019ApJ...886...79G,2020PASJ...72...60S,2020MNRAS.494.1693K}. The typical host DMH mass can be studied by the large-scale ($>1-2\,{\rm Mpc}$) bias determined by the ACF of AGNs or the CCF between AGNs and galaxies, assuming the linear-biasing scheme. 
One of the popular ways of interpreting the correlation function (CF) is the halo occupation distribution (HOD) analysis \citep[e.g.,][]{2000MNRAS.318.1144P,2000MNRAS.318..203S,2002PhR...372....1C}
, where the CF is modeled as the sum of one-halo (1-h) and two-halo (2-h) terms, representing the contributions of pairs that are within the same DMHs and across different DMHs respectively. By the HOD analysis, one can obtain, not only the "typical" DMH mass but also constraints on how the objects are distributed among the DMHs as a function of the DMH mass ($M_{\rm DMH}$). The tool developed for the HOD analysis is also used to measure the bias parameter accurately by applying it to the linear regime or applying it to all scales. 
This has often been useful to improve on the linear bias measurements, where previous studies used the power-law approximation of CFs. The linear regime of the CFs is the Fourier transform of the mass density linear power spectrum and not a power-law. Thus using the proper linear power spectrum included in the HOD tools is a large improvement than using the simple power-law model, even in the case where information from the 1-halo term is not used, e.g., due to the lack of sufficient signal at small scales, as is often the case for AGN ACFs. 

X-ray-selected AGN clustering studies have been made mainly through ACFs \citep{2007ApJS..172..396M, 2008ApJS..179..124U,2009A&A...494...33G,2011ApJ...741...15S,2013MNRAS.428.1382K,2014ApJ...796....4A,2019A&A...632A..88A,2019A&A...629A..14V}. On the other hand, due to the small number of AGN samples, taking a cross-correlation approach with a much larger sample of galaxies would give AGN clustering measurements with better statistical accuracy \citep{2011ApJ...726...83M,2012ApJ...746....1K,2018MNRAS.474.1773K}. Also, the limited spatial resolution of X-ray images is less of a problem with the CCF approach than the ACF approach, since we use the positions of the optical counterparts selected by, e.g., the likelihood ratio technique \citep{2011ApJ...742...61S,2016ApJ...817...34M, Miyaji2024} or a Bayasian-based statistics \citep[e.g.][]{Salvato2018}, and measure separations to the galaxies. While the spatial resolution is less of a problem with {\it Chandra}, its point spread function (PSF) degrades with off-axis angle rapidly and the sources detected at large off-axis angles still suffer from the spatial resolution problem. This is a problem with an HOD-type analysis by hampering the 1-halo term measurements probed in small scales.

Motivated by the above situation, precise clustering measurements through CCF studies and their HOD modelings of X-ray-selected AGNs have been made. The HOD modeling of X-ray-selected AGNs up to $z\sim 1$ (\citealt{2011ApJ...726...83M,2012ApJ...746....1K,2018MNRAS.474.1773K}). \cite{2011ApJ...726...83M} performed the HOD modeling of the CCF at $z<1$ and found that satellite slope, $\alpha_{s}<1$ (see Eq. 14). At $z>1$, the HOD modeling of the CCF has not yet been investigated; therefore, the small-scale clustering at $z>1$ has not yet been studied. Motivated by the situation described above, we perform the HOD Modeling of the CCF of galaxies and X-ray-selected AGNs at $0.6<z<2.6$ in the COSMOS field to present a precise AGN clustering at $z\sim1$ and $\sim2$ in the Universe.  

The outline of this paper is as follows.
In Section 2, we describe the data that we use in this study and the selection criteria for galaxies and AGNs.
In Section 3, we describe the methods for the clustering measurements. 
In Sections 4, 5, and 6, we give our results, discussion, and summary. In Appendix \ref{sec:app_covariance}, we explain a problem we have encountered with our covariance matrix and its remedy. Throughout this paper we adopt a $\Lambda$CDM cosmology with $\Omega_m$ = 0.3,\  $\Omega_{\Lambda}$ = 0.7, $\sigma_8(z=0)=0.8$, and a Hubble constant of $H_0=70\,h_{\rm 70} {\rm km\,s^{-1}Mpc^{-1}}$ for X-ray luminosities and stellar masses, while $H_0 = 100\,h\, {\rm km\,s^{-1} Mpc^{-1}}$ is used for distances to aid in better comparison with previous works. 
The AB magnitude system is used (\citealt{1974ApJS...27...21O}). 
\begin{table*}[!hbt]

\begin{center}
\caption{Properties of galaxy samples\label{tab:gal_props}}
\begin{tabular}{ccc@{\hspace{0.2cm}}c@{\hspace{0.2cm}}c@{\hspace{0.2cm}}c@{\hspace{0.2cm}}c@{\hspace{0.2cm}}c@{\hspace{0.2cm}}c@{\hspace{0.2cm}}c@{\hspace{0.2cm}}c} \hline \hline
         Sample Name & Redshift Range& $\log M_{*}$ Range & Number & $\langle n_{\rm gal}\rangle$ & $\langle z_{\rm med}\rangle$ & $\langle \log M_{\rm *}\rangle$& \\
        &&($h^{-1}\,M_{\odot}$)  && ($h^{3}$ Mpc$^{-3}$) & & ($h^{-1}\,M_{\odot}$)  \\ \hline
 $z\sim1$ galaxy sample &   $0.6<z<1.4$ &$9.5<\rm log  (\it M_{*}/M_{\odot})< \rm 10.0$ & 22,281 & $8.40\times10^{-3}$ & $1.01$& 9.72&\\
  $z\sim2$ galaxy sample &   $1.4<z<2.6$ &$9.5< \rm log (\it M_{*}/M_{\odot})< \rm 10.0$ & 18,874 & $3.22\times10^{-3}$ & $1.75$&9.73 &\\

           \hline \\
                        \end{tabular}

\end{center}
\end{table*}

\begin{table*}[!hbt]

\begin{center}
\caption{Properties of X-ray-selected AGN samples\label{tab:xraysamp}}

\begin{tabular}{ccc@{\hspace{0.2cm}}c@{\hspace{0.2cm}}c@{\hspace{0.2cm}}c@{\hspace{0.2cm}}c@{\hspace{0.2cm}}c@{\hspace{0.2cm}}c@{\hspace{0.2cm}}c@{\hspace{0.2cm}}c} \hline \hline
         Sample Name &   Redshift Range& $\log L_{\rm X}$ Range & Number & $\langle n_{\rm AGN}\rangle$ & $\langle z_{\rm med}\rangle$ & $\langle \log L_{\rm X}\rangle$ \\
        && ($h_{70}^{-2}\ \rm erg\,s^{\rm -1}$) && ($h^{3}$ Mpc$^{-3}$) & &($h_{70}^{-2}\,\rm erg\,s^{\rm -1}$) \\ \hline
        
         \hline
 $z\sim1$ X-ray-selected AGN sample &   $0.6<z<1.4$ &log $L_{X}>42$ & 718 &$2.71\times10^{-4}$  & $1.02$& 42.71& \\
  $z\sim2$ X-ray-selected AGN sample &   $1.4<z<2.6$ & log $L_{X}>42$ & 626 & $1.23\times10^{-4}$ &$1.80$&43.2 &\\
           \hline \\
                        \end{tabular}

\end{center}

\end{table*}

\begin{figure*}[!t]
\begin{center}
\includegraphics[bb= 0 0 820 330,clip,width=18.3cm]{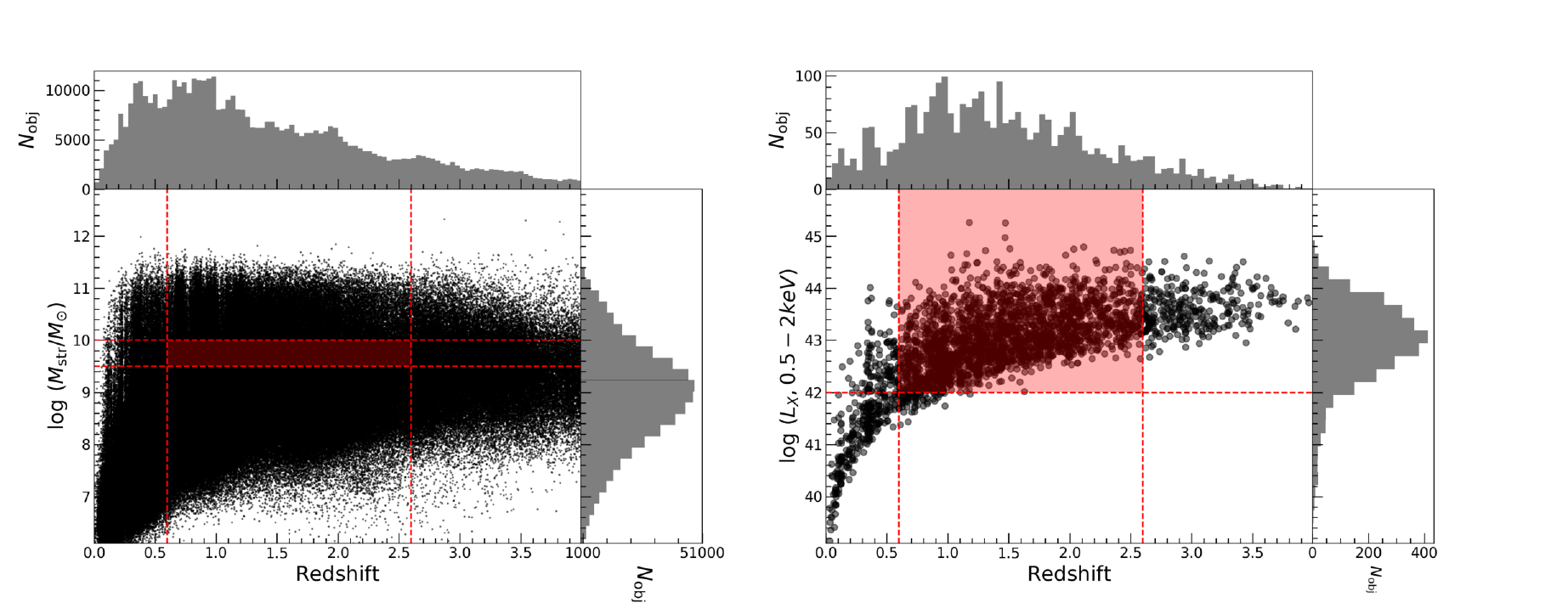}
\caption{Left: Distribution of stellar mass as a function of redshift of galaxies at $0<z<4$ in COSMOS2015 catalog. The red-shaded region denotes the selection criteria of our sample. Right: Distribution of $0.5-2$ keV X-ray luminosity as a function of redshift of {\sl Chandra} legacy X-ray-selected AGNs at $0<z<4$. Red shaded region denotes the selection criteria of our sample. \label{fig:z_vs_mstar}
}  
\end{center}
\end{figure*}
\section{Data}

\subsection{Galaxy Sample at $0.6<z<2.6$}
We construct volume-limited stellar mass-selected samples of galaxies by utilizing the COSMOS2015 catalog (\citealt{2016ApJS..224...24L}\footnote{Recently, the COSMOS2020 catalog (\citealt{2022ApJS..258...11W}) has become public. Most of the work in this paper has been made before the publication of the COSMOS2020 catalog.}) to calculate the ACF of galaxies and CCF of galaxies and X-ray-selected AGNs.
This catalog has $>30$ band photometry and accurate photometric redshift ($z_{\rm ph}$) information over the 2 square degrees of COSMOS. The photometric redshift precision of $\sigma_{\Delta z/(1+z_{\rm sp})} = 0.007$. The stellar mass-redshift plane of galaxies is shown in the left panel of Figure \ref{fig:z_vs_mstar}. We try to make the galaxy sample as homogeneous and volume-limited as possible to minimize the effects of redshift evolution within a redshift bin. 

In order to construct volume-limited stellar mass-selected samples of galaxies, we utilize two selection criteria as follows:
\begin{eqnarray}
9.5< \rm log \it ( M_{*} / M_{\odot}) <\rm 10.0,\\
K_{AB}<25.0,
\label{eq:mass_k_cut}
\end{eqnarray}
where, $M_{*}$ is the stellar mass of galaxies.  
In order to study the redshift evolution, we divide our galaxy sample in two bins of redshift.
To construct $z\sim1$ galaxy sample, 
we add the selection criteria about the redshift of galaxies with known spectroscopic redshifts:
\begin{equation}
0.6<z_{\rm sp}<1.4.
\label{eq:z1range}
\end{equation}
We include objects without spectroscopic redshift measurements if they have photometric redshift measurements and non-negligible probabilities of being within this redshift range. Specifically, we include an object if its photometric redshift probability distribution function ($\mathrm{PDF\it z}(z)$), provided with the COSMOS2015 catalog satisfies:
\begin{equation}
\mathrm{PDF\it z}(0.6<z_{\rm ph}<1.4) > 0.025.
\label{eq:z1qpdf}
\end{equation}
To construct $z\sim2$ galaxy sample, we add the selection criteria about the redshift of galaxies with known spectroscopic redshifts:
\begin{equation}
1.4<z_{\rm sp}<2.6.
\label{eq:z2range}
\end{equation}
For those without known spectroscopic redshifts but with photometric redshift measurements, we include those that satisfy:
\begin{equation}
\mathrm{PDF\it z}(1.4<z_{\rm ph}<2.6) > 0.025.
\label{eq:z2pdf}
\end{equation}
Using Eqs. 1--6, we construct $z\sim1$ and $z\sim2$ stellar mass-selected galaxy samples. These galaxy samples are listed in Table \ref{tab:gal_props} and the spatial and redshift distribution of these samples are shown in Figure 2.
\begin{figure*}[!t]
\begin{center}
\includegraphics[bb= 0 0 730 700,clip,width=15.5cm]{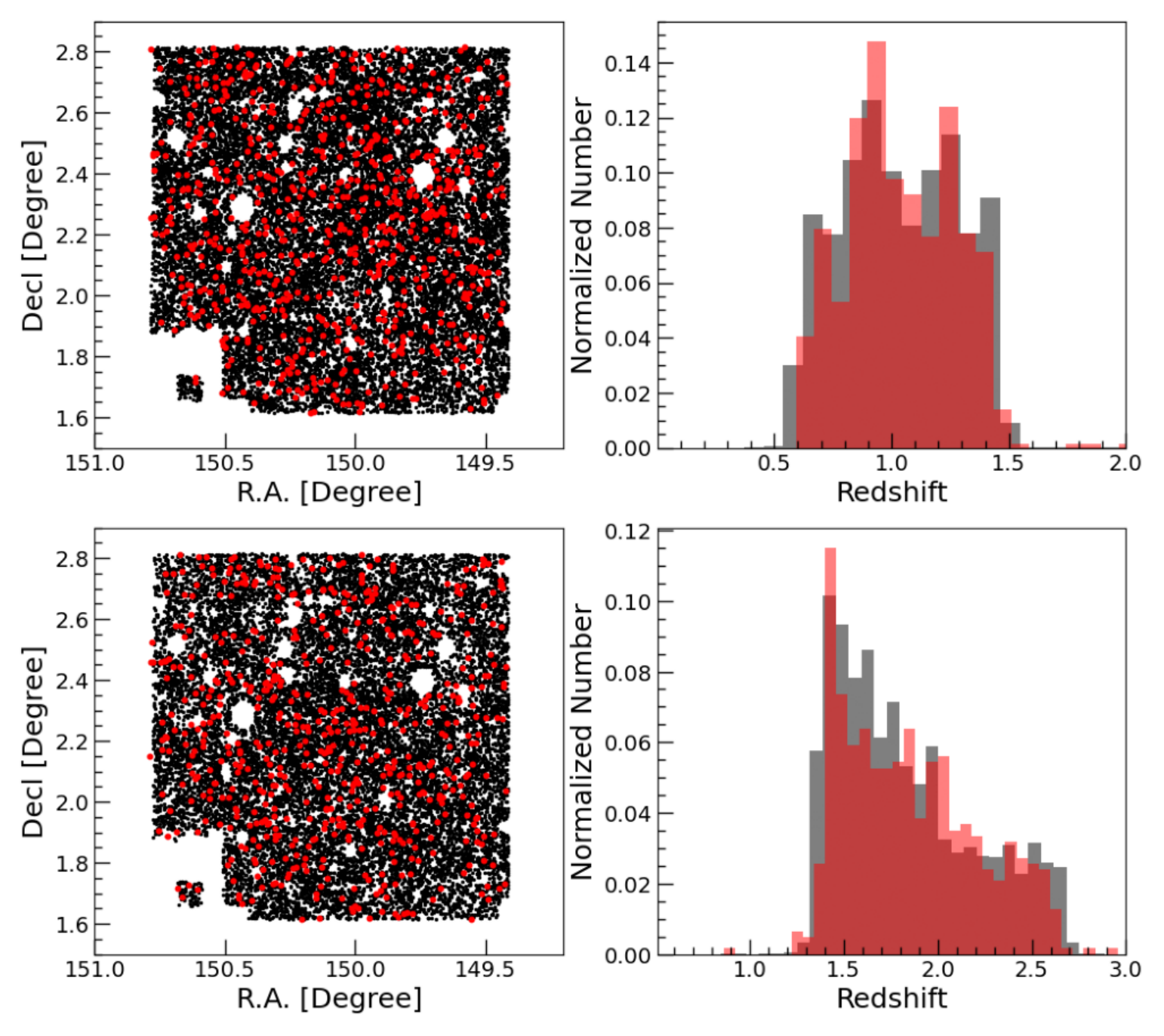}
\caption{Upper left panel: The spatial distributions of the X-ray-selected AGNs and galaxies for the $0.6<z<1.4$ samples. Red and black circles represent X-ray-selected AGNs and galaxies, respectively. Upper right panel: The redshift distributions of the X-ray-selected AGNs and galaxies for the $0.6<z<1.4$ samples. Red and black histograms represent the X-ray-selected AGN and galaxy samples, respectively. Lower panels: The same as the upper panels but for the X-ray-selected AGNs and galaxies in the $1.4<z<2.6$ samples.\label{fig:spatial_z_dist}}  
\end{center}
\end{figure*}

\begin{figure*}[!t]
\begin{center}
\includegraphics[bb= 0 0 700 680,clip,width=19cm]{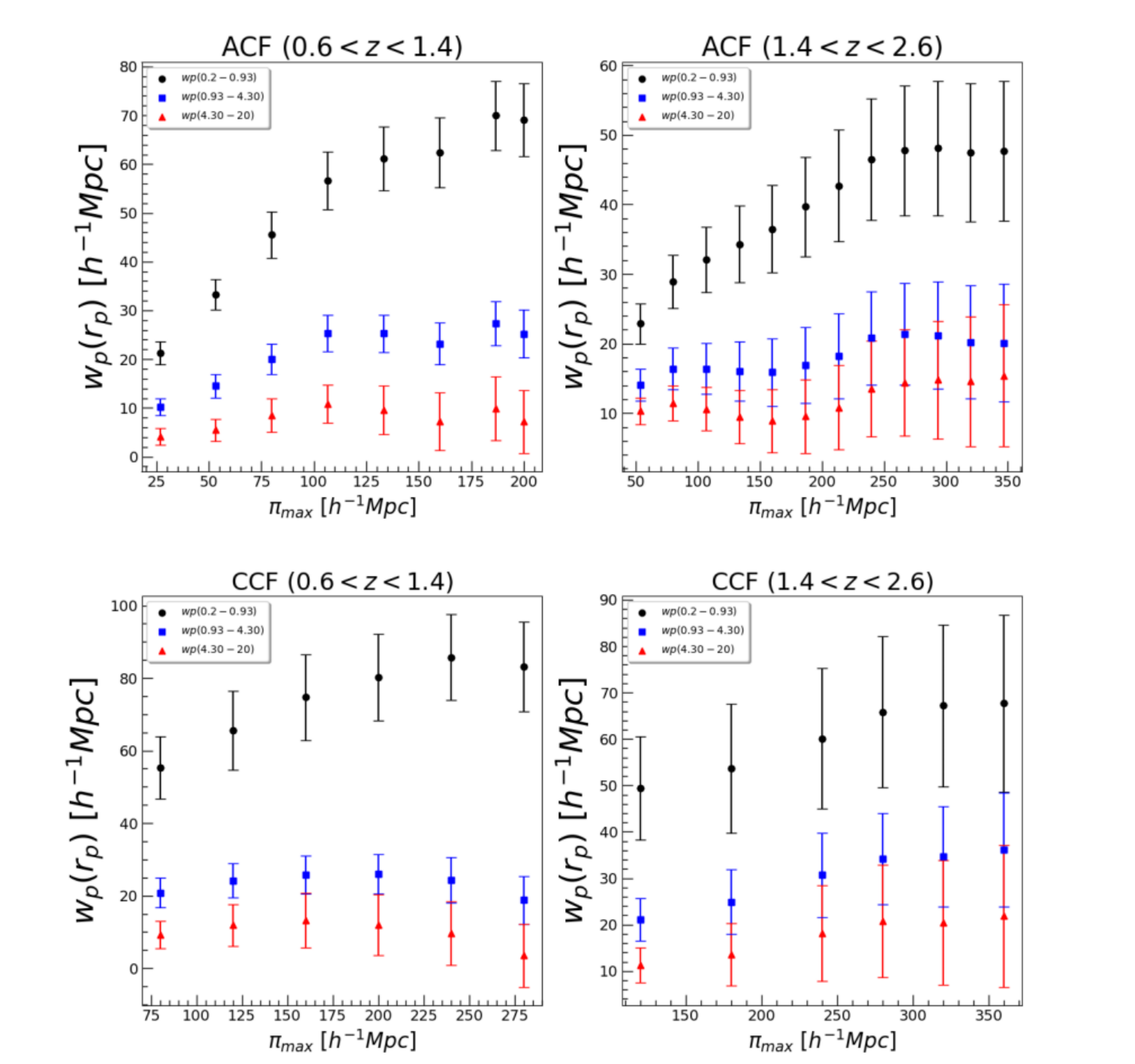}
\caption{Upper panels: The ACF of galaxies at $0.6<z<1.4$ (left) and $1.4<z<2.6$ (right) as a function of $\pi_{\rm max}\,[h^{-1}{\rm Mpc}]$. Black circles, blue squares, and red triangles are ACF at $0.2 < r_{\rm p}\,[h^{-1}{\rm Mpc}] < 0.93$,  $0.93 < r_{\rm p}\,[h^{-1}{\rm Mpc}] < 4.30$, and  $4.30 < r_{\rm p}\,[h^{-1}{\rm Mpc}]< 20.0$, respectively. Lower panels: Same as the upper panel but for the CCF of galaxies and AGNs.
\label{fig:wp_vs_pimax}}
\end{center}

\end{figure*}

\subsection{X-ray-selected AGN Sample at $0.6<z<2.6$}
We construct an X-ray-selected AGN sample by utilizing the {\sl Chandra} Legacy Identification Catalog (\citealt{2016ApJ...817...34M}) to calculate the CCF of galaxies and X-ray-selected AGNs. This catalog contains the optical and infrared counterparts of the X-ray sources for the X-ray sources detected in the {\sl Chandra} COSMOS-Legacy Survey (\citealt{2016ApJ...819...62C}). This catalog also includes information of the spectroscopic and photometric redshift. The photometric redshifts in \citet{2016ApJ...817...34M} are derived independently from those in \citet{2016ApJS..224...24L} and optimized for X-ray-selected AGNs following \citet{2011ApJ...742...61S}. Those with photometric redshifts only are included in the sample in the same way as the galaxy sample using the $\mathrm{PDFz}$'s (denoted as $\mathrm{PDz}$ in \citealp{2016ApJ...817...34M}). We consider X-ray sources that have 
$L_{\rm X}>10^{42}\, h_{\rm 70}^{-2}\, {\rm erg\,s^{-1}}$ ($0.5-2$ keV) secure AGNs and include in our sample. Given the limited number of X-ray-selected AGNs, constructing a volume-limited sample of AGNs is not feasible, and thus our AGN sample is essentially flux-limited. The $\log\, L_{ \rm X}(0.5-2 \rm keV)$-redshift plane of X-ray-selected AGN is shown in the right panel of Figure \ref{fig:z_vs_mstar}.

We choose the $z\sim 1$ and $z\sim 2$ X-ray-selected AGN samples in the same way as the corresponding galaxy 
samples, following Eqs.~\ref{eq:z1range}-\ref{eq:z2pdf}. 

These samples are listed in Table \ref{tab:xraysamp} and the spatial and redshift distribution of these samples are shown in Figure 2.


\begin{figure*}[!t]
\begin{center}
\includegraphics[bb= 0 120 650 700,clip,width=19cm]{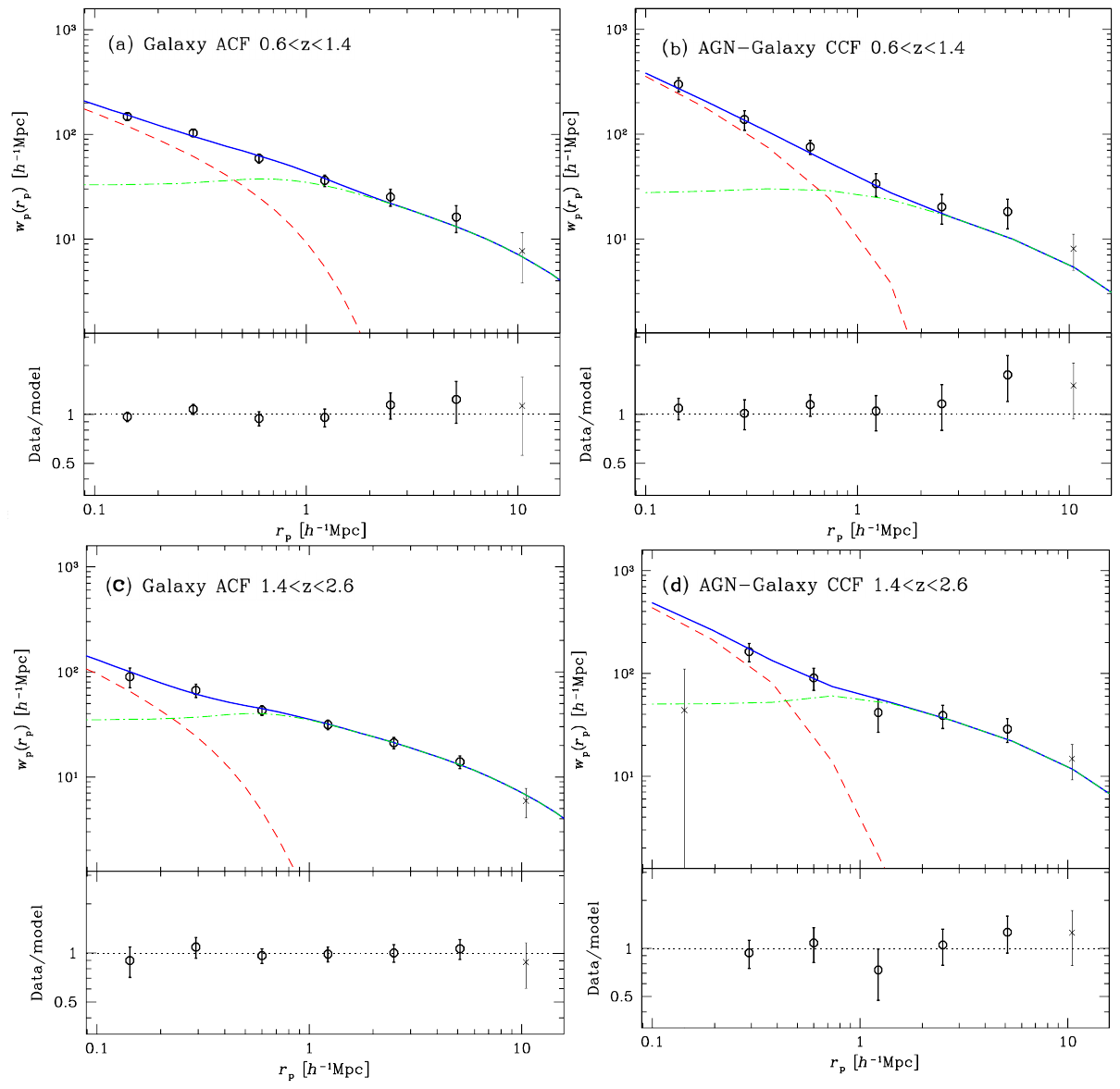}
\caption{Upper left panel: The ACF of galaxies at $0.6<z<1.4$, and the best-fit HOD model with residuals. The data points and error bars show our measurements of the galaxy ACF. The circles show the data points that are used for the fit, while the cross shows those excluded. The short-dashed
(red), long-dashed (green), and solid (blue) lines show the 1-halo term, 2-halo term, and the total, respectively Upper right panel: Same as the upper left panel but for the CCF of galaxies and AGNs at $0.6<z<1.4$. Lower left panel: Same as the upper left panel but for the ACF of galaxies at $1.4<z<2.6$. Lower right panel: Same as the upper left panel but for the CCF of galaxies and AGNs at $1.4<z<2.6$.\label{fig:wprp_all}}  
\end{center}
\end{figure*}

\section{Methods}

\subsection{ACF of galaxies}
\label{sec:acf}
In order to calculate ACFs of galaxies,  we use the \cite{1993ApJ...412...64L} estimator, in a two-dimensional grid in the $(r_{\rm p},\pi)$ space, where $r_{\rm p}$ is the projected distance and $\pi$ is the line-of-sight separation:
\begin{equation}
\xi_{\rm ACF}(r_{\rm p},\pi)=\frac{DD(r_{\rm p},\pi) -2 DR(r_{\rm p},\pi) + RR(r_{\rm p},\pi)}{RR(r_{\rm p},\pi)},
\end{equation}
where, $DD$, $DR$, and $RR$ are the normalized number of pairs \citep[see e.g. Eq. (3) of][]{2007ApJS..172..396M} within the real sample, the number of pairs between the real sample and random sample, and the number of pairs within the random sample, respectively. The random sample is created with 100 times the number of our galaxy samples to match the redshift distribution of our galaxy sample.
When an object has a photometric redshift only, it is included as if it were multiple objects at the same sky position with different redshifts weighted by its $\mathrm{PDFz}(z)$ within the redshift range of interest, following the approach of \citet{2016ApJ...832...70A}.

 We then calculate the projected correlation function:
\begin{equation}
w_{\rm p,ACF}(r_{\rm p})=2 \int_0^{\pi_{\rm max}}\xi_{\rm ACF}(r_{\rm p},\pi) \textrm{d}\pi,
\label{eq:wprpint}
\end{equation}
where, $\pi_{\rm max}$ is determined by the saturation of the integral. 
We calculate the errors of ACF by using the jackknife resampling technique. The error and covariance matrix estimations are explained in Sect. 3.3. 
The saturating $\pi_{\rm max}$ value is found by plotting the projected galaxy auto-correlation function values in three large $r_{\rm p}$ bins as a function of $\pi_{\rm max}$. Figure \ref{fig:wp_vs_pimax} (upper panels) shows that the amplitude of the projected galaxy auto-correlation function saturates at $\pi_{\rm max}\sim 190 h^{-1}$ Mpc for  $0.6<z<1.4$ and $\pi_{\rm max}\sim 240 h^{-1}$ Mpc for $1.4<z<2.6$, respectively. Thus we adopt these $\pi_{\rm max}$ values for the ACFs for our subsequent analysis.

\begin{figure*}[!hbt]
\begin{center}
\includegraphics[bb= 0 50 630 810,clip,width=16.5cm]{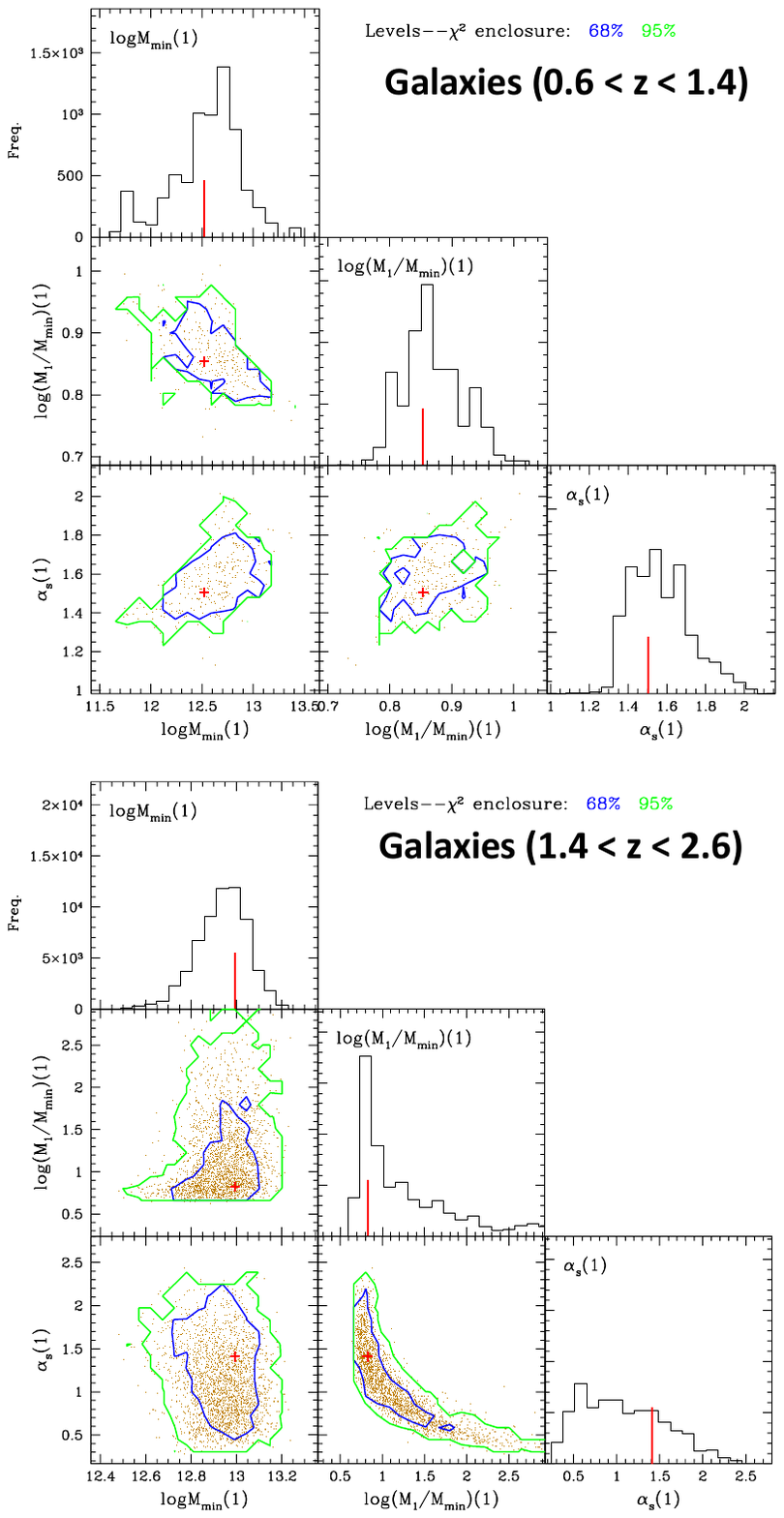}
\caption{Upper panel: The confidence contours of the best-fit HOD model for galaxies at $0.6<z<1.4$. These confidence contours are estimated from the joint fit to the galaxy ACF and the galaxy-AGN CCF for the $0.6<z<1.4$ sample. The red crosses in each off-diagonal panel denote the best-fit values in the corresponding two parameters, while the red vertical line in each diagonal panel is the best-fit value of the corresponding one parameter.  Blue and green contours are drawn at $\chi^2$ levels below which 68\% and 95\% of the MCMC chain points fall respectively. The small brown dots represent the MCMC points. Each diagonal panel shows the histogram of the distributions MCMC points in one parameter space as labeled. Each parameter name label is accompanied by a '(1)' to indicate that it is for the galaxy sample.   
Lower panel: Same as the upper panel but for the $1.4<z<2.6$ galaxy sample.\label{fig:gal_hod_contours}}  
\end{center}

\end{figure*}

\begin{figure*}[!ht]
\begin{center}
\includegraphics[bb= 0 0 630 820,clip,width=16cm]{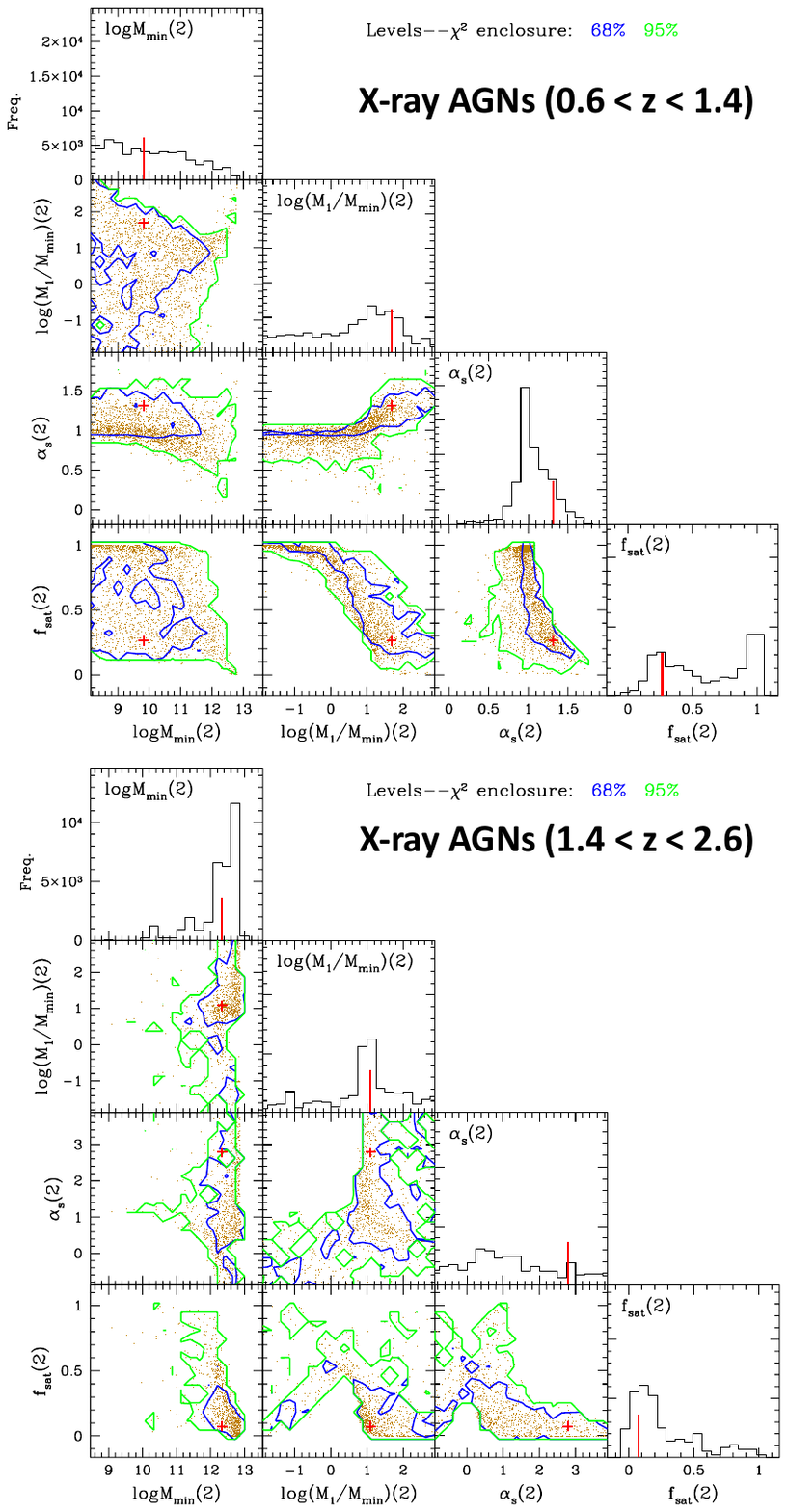}
\caption{Upper panel: The confidence contours of the best-fit HOD model for the $0.6<z<1.4$ X-ray-selected AGNs sample. These confidence contours are obtained from the joint fit to the ACF and galaxy-AGN CCF. The meanings of symbols and line colors are the same as those of Figure 5. The attachment '(2)' in each parameter name panel indicates that it is for the X-ray-selected AGN sample. Lower panel: Same as the upper panel but for the $1.4<z<2.6$ X-ray-selected AGN sample.
\hspace{3cm} \label{fig:agn_hod_contours}  }
\end{center}

\end{figure*}



\begin{table*}[!hbt]
\caption{The HOD Parameters\label{tab:hod_params}}
\begin{center}
\begin{tabular}
{@{\hspace{0cm}}c@{\hspace{0.1cm}}c@{\hspace{0.1cm}}c@{\hspace{0.1cm}}c@{\hspace{0.1cm}}c@{\hspace{0.1cm}}c@{\hspace{0.1cm}}c@{\hspace{0.1cm}}c@{\hspace{0.1cm}}c@{\hspace{0.1cm}}c@{\hspace{0.1cm}}c@{\hspace{0.1cm}}c@{\hspace{0.1cm}}} \hline \hline
         Sample  &   Redshift Range & $\log M_{\rm min}$  & $\log \langle M_{\rm DMH}\rangle$ & $\alpha_{s}$ & linear bias & $\log M_{\rm typ}$ & $\log(M_1/M_{\rm min})$\\
          & &($h^{-1}\,M_{\odot}$) & ($h^{-1}\,M_{\odot}$)  & &  & ($h^{-1}\,M_{\odot}$) \\
                  \hline
         galaxy &   $0.6<z<1.4$  & $12.89  \ (12.31;12.95)$ &  $12.85 \ (12.78;12.86)$        & $1.58\ (1.42;1.74)$      &        $1.50\ (1.33;1.52)$    & $12.42\ (12.16;12.44)$               &   $0.82\ (0.81;0.93)$                     \\
          galaxy &   $1.4<z<2.6$  & $12.99 \ (12.83;13.04)$   &    $12.45\ (12.37;12.46)$      &  $1.41\ (0.57;1.62)$           & $2.03\ (1.92;2.06)$   & $12.18\ (12.06;12.20)$               &   $0.82\ (0.78;1.03)$                      \\
 AGN &   $0.6<z<1.4$  & $9.82 \ (8.54;11.37)$ &    $12.84\ (12.72;12.92)$       &  $1.31\ (0.90;1.29)$         & $1.16\ (1.16;1.31)$ & $11.82\ (11.82;12.12)$  &   $1.69\ (-0.93;1.82)$                                     \\
 AGN  &  $1.4<z<2.6$ & $12.34\ (11.67;12.78)$  &$13.01\ (12.61;13.13)$       & $2.79\ (-0.13;2.77)$          & $2.95\ (2.30;3.55)$  & $12.80\ (12.38;13.06)$  &   $1.09\ (-0.45;1.99)$  \\
           \hline
\end{tabular}
\end{center}
\end{table*}

\subsection{CCF between galaxies and AGNs}
\label{sec:ccf}
To calculate the CCFs of galaxies and AGNs, we use \cite{1993ApJ...412...64L} estimator which is modified for the CCF (\citealt{2003ApJ...584...45A,2007ApJ...663..765K,2011MNRAS.415.2626B}),
\begin{eqnarray}
 \nonumber &\xi_{\rm CCF}(r_{\rm p},\pi)=
&\frac{D_{\rm G}D_{\rm A}-D_{\rm G}R_{\rm A}-R_{\rm G}D_{\rm A}+ R_{\rm G}R_{\rm A}}{R_{\rm G}R_{\rm A}}\\
\label{eq:ccfest}
\end{eqnarray}
where $D_{\rm G}D_{\rm A}$, $D_{\rm G}R_{\rm A}$, $R_{\rm G}D_{\rm A}$ and $R_{\rm G}R_{\rm A}$ are normalized
numbers of galaxy-AGN, galaxy-randomized AGN, randomized galaxy-AGN, and randomized galaxy-randomized AGN pairs falling into the given $(r_{\rm p},\pi)$ bin respectively.
The random sample for galaxies and AGNs is created with 100 times the number of our galaxy and AGN sample to match the redshift distribution of them.
The $\mathrm{PDF\it z}$'s of the photometric redshifts in the galaxy and AGN samples are treated in the same way as the galaxy ACF case.

 We then calculate the projected correlation function:
\begin{equation}
w_{\rm p,CCF}(r_{\rm p})=2 \int_0^{\pi_{\rm max}}\xi_{\rm CCF}(r_{\rm p},\pi) \textrm{d}\pi,
\end{equation}

where, $\pi_{\rm max}$ is determined by the saturation of the integral. 
The lower panels of Figure \ref{fig:wp_vs_pimax} shows $w_{\rm p}$ estimates for the CCF in three $r_{\rm p}$ bins as functions of $\pi_{\rm max}$ in the same way as the ACF. From this figure, we observe that the projected CCF estimates saturate at  $\pi_{\rm max}\sim 200 h^{-1}$ Mpc for $0.6<z<1.4$ and at $\pi_{\rm max}\sim 300 h^{-1}$ Mpc for $1.4<z<2.6$, respectively. We adopt these $\pi_{\rm max}$ values for the CCFs for our subsequent analysis.

\subsection{Errors and Covariance Matrix}
\label{sec:error_cov}
We calculate the errors of ACF and CCF by using the jackknife resampling technique.
We divide the survey area into $N=16$ ($4 \times 4$) subsections. 
The $k$-th ($k=1..N$) jackknifed sample is constructed by excluding 
the objects (both from the real and random samples) in the $k$-th subsection and 
$w_{{\rm p}ik}$ is the ACF or CCF, depending on the $i$ value, calculated from the $k$-th jackknifed sample. The index $i$ runs through $r_{\rm p}$ bins of the ACF followed by those of CCF. The $N$ jackknife-resampled correlation functions define the covariance matrix, $M_{ij}$: 
\begin{eqnarray}
 M_{ij} &=& \frac{N -1}{N} \sum_{k=1}^{N} \left[w_{{\rm p}ik}(r_{{\rm p},i})-\langle w_{{\rm p}i}
  (r_{{\rm p},i})\rangle \right]\nonumber\\
   &\times& \left[w_{{\rm p}jk}(r_{{\rm p},j})-\langle w_{{\rm p}j}(r_{{\rm p,j}})\rangle \right]. 
   \label{eq:cov}
\end{eqnarray}
The 1$\sigma$ error bar for the $i$-th bin corresponds to $\sqrt{M_{ii}}$.
In constructing the covariance matrix, we initially considered not only the errors in different $r_{\rm p}$ bins within each of the ACF and CCF but also those across the ACF and CCF are correlated. However, due to inaccuracies of the off-diagonal elements, especially those across ACF/CCF, we only use the off-diagonal elements within the CCF, as explained in Appendix \ref{sec:app_covariance}.

\subsection{The HOD modeling}
We perform the HOD modeling, following an approach similar to \cite{2011ApJ...726...83M}. 
We model the two-point projected correlation function as the sum of two terms,
\begin{equation}
  w_{\rm p}(r_{\rm p})=w_{\rm p,1h}(r_{\rm p}) + w_{\rm p,2h}(r_{\rm p}),
\end{equation}
where $w_{\rm p,1h}(r_{\rm p}) $ and $w_{\rm p,2h}(r_{\rm p}) $ are the 1-halo term contributed by pairs where both of objects occupy the same DMH and the 2-halo term contributed by pairs where the two objects occupy different DMHs, respectively. We use the halo mass function of \cite{2001MNRAS.323....1S}, the Navarro-Frenk-White (NFW) dark matter halo profile (\citealt{1997ApJ...490..493N}) for the DMH density distribution in our HOD modeling. By performing a correlated $\chi^2$ fit to both the ACF of galaxies and galaxy -- AGN CCF jointly, we can constrain the HOD of X-ray-selected AGNs.

Note that there have been several improvements and modifications of our HOD code after \cite{2011ApJ...726...83M}. The scale-dependent bias and effects of halo-halo collision formulated by \citet{2005ApJ...631...41T} have been included since the version of the code used in \citet{2018MNRAS.474.1773K}. The Markov-Chain Monte Carlo (MCMC) method for the HOD parameter search has also been included and used in \citet{krumpe23}, where a two-step process was taken. In the first step, galaxy HOD was determined from the galaxy ACF and, in the second step, the AGN HOD was determined from the galaxy-AGN CCF assuming the fixed best-fit galaxy HOD parameters obtained in the first step, with separate MCMC chains. In this work, we search for HOD parameters and confidence ranges for the galaxies and AGNs by simultaneous MCMC parameter search over the galaxy ACF and galaxy-AGN CCF.

\subsubsection{The parameterized HOD model of galaxies}
One of the most common forms of the galaxy HOD is a combination of a step function with a smooth transition for the galaxies at halo centers and a power-law form multiplied by the central HOD, proposed by \cite{Zheng_2005}.
This form is intended for luminosity-thresholding samples of galaxies, where it is assumed that the centers of 
most massive halos are always occupied by a galaxy in the sample. Since our galaxy sample is not luminosity-thresholding 
sample, the central HOD at the highest halo mass end does not have to saturate at unity. Since the HOD predicted correlation function does not depend on the absolute normalization of the HOD, we express central HOD with proportionality rather than equality. In comparing the HOD-model predicted correlation function with the observed one, we also do not put the density constraint, since such constraint is meaningful when the central HOD saturates at unity, which is not the case for our sample. Thus our galaxy HOD has the form (\citealt{2007ApJ...667..760Z}):

\begin{eqnarray}
\langle N_{\rm G,c}\rangle (M_{\rm DMH}) &\propto & 
\frac{1}{2}\left[1+\mathrm{erf}\left(\frac{\log M_{\rm DMH}-\log M_{\rm
min}}{\sigma_{\log M}}\right)\right]\nonumber ,\\
\end{eqnarray}

for the centrals and the following form for the satellites:

\begin{eqnarray}
\langle N_{\rm G,s}\rangle (M_{\rm DMH}) &=& \langle N_{\rm G,c}\rangle (M_{\rm DMH})\;
\left(\frac{M_{\rm DMH}}{M_1}\right)^{\alpha_{\rm s}}.
 \end{eqnarray}
 
 Where $M_{\rm min}$, $\sigma_{\rm log \it M}$, and $M_{1}$ are the characteristic minimum
mass of halos that can host central galaxies, the width of the cutoff profile, and the normalization of the satellite HOD, respectively.

\begin{figure*}[!t]
\begin{center}
\includegraphics[width=15cm]{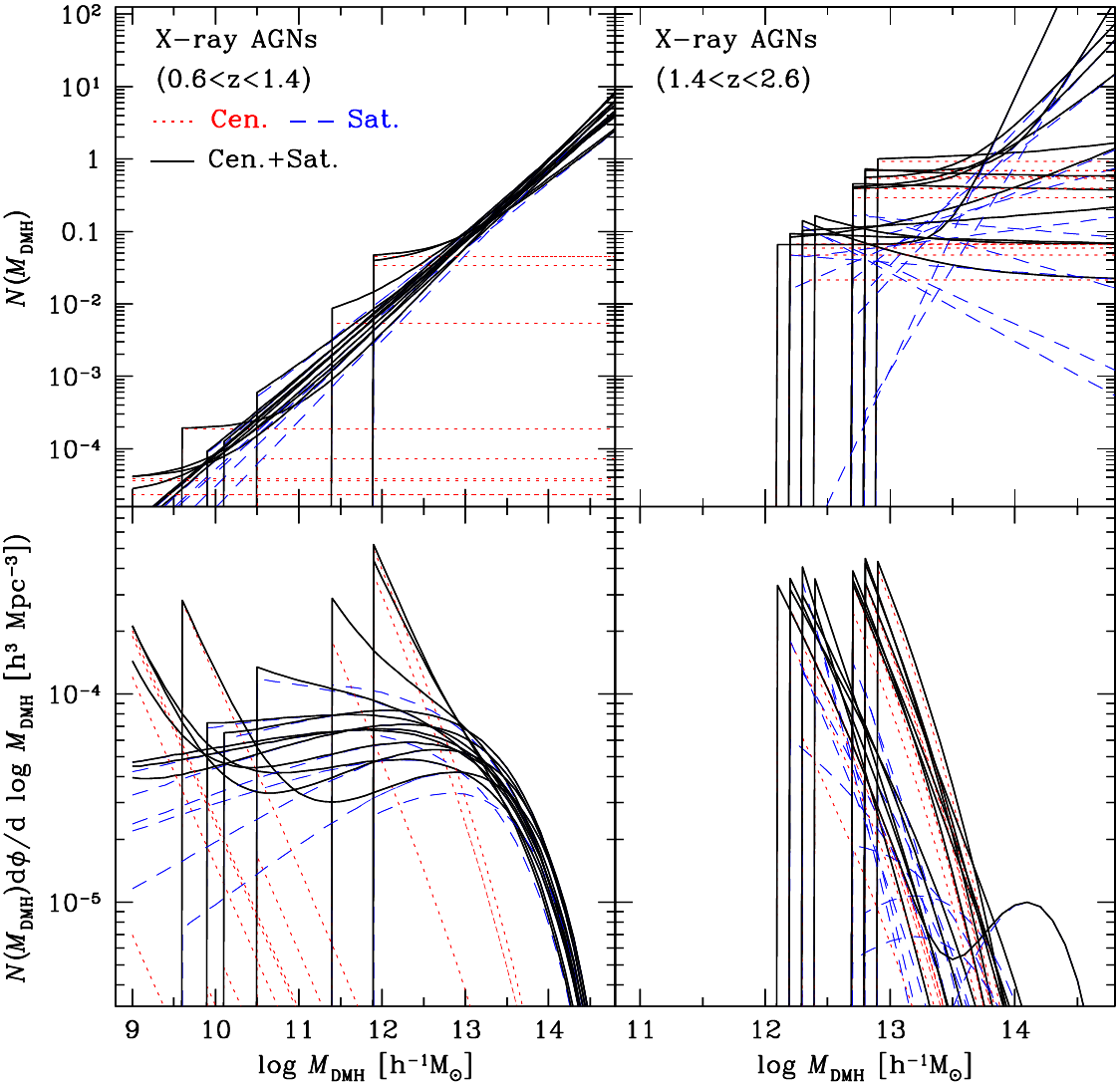}
\caption{Halo occupation distributions of accepted models for the X-ray-selected AGN sample. The upper left and right panels show the HODs for the $0.6<z<1.4$ and $1.4<z<2.6$ AGN samples respectively. In each panel, the red dotted, blue dashed, and black solid lines show acceptable randomly selected models for central ($\langle N_{\rm A,c} \rangle$), satellite  ($\langle N_{\rm A,s} \rangle$) and the central+satellite HODs. The models are selected from chain points of our MCMC run that are among 68\% smallest $\chi^2$ of all chain points. 
The lower panels show the central+satellite HODs multiplied by the DMH mass function.
\label{fig:agn_hod_plots}
}
\end{center}
\end{figure*}

\begin{figure*}[!t]
\begin{center}
\includegraphics[bb= 0 0 800 480,clip,width=15.2cm]{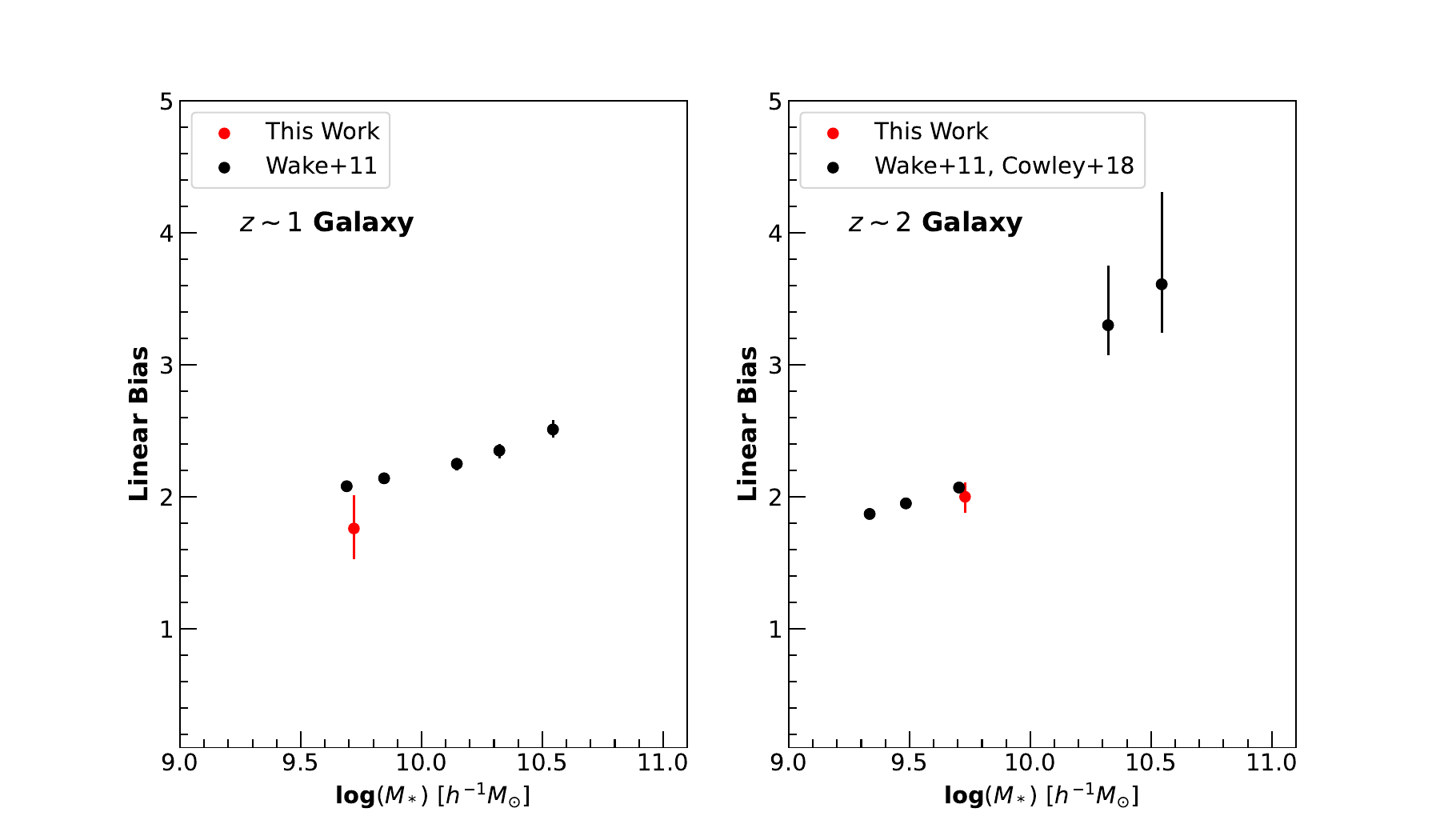}
\caption{Left:$z\sim1$ Galaxy linear bias as a function of the stellar mass. Red and black points represent our results and results in the literature (\citealt{2011ApJ...728...46W}). Error bars of our results are at the $68\% $ confidence level (between $16\% $ and $84\% $ percentiles from the corresponding MCMC chains) for one interesting parameter. Right: Same as the left panel but for $z\sim2$ galaxy linear bias. Red and black points represent our results and results in the literature (\citealt{2011ApJ...728...46W,2018ApJ...853...69C}). \label{fig:gal_bias_mstar}}  
\end{center}

\end{figure*}
\subsubsection{The parameterized HOD model of X-ray-selected AGNs}
In our HOD model of X-ray-selected AGNs, we use the occupation function for central AGNs as follows,

\begin{eqnarray}
\langle N_{\rm A,c}\rangle &=& f_{\rm A}\Theta(M_{\rm DMH}-M_{\rm min}), 
\label{eq:agnhodcen}
 \end{eqnarray}
and the occupation function for satellite AGNs is expressed as follows,
\begin{eqnarray}
\langle N_{\rm A,s}\rangle &=& f_{\rm A}\Theta(M_{\rm DMH}-M_{\rm min})(M_{\rm DMH}/M_1)^{\alpha_{\rm s}}, 
\label{eq:agnhodsat}
\end{eqnarray}

where $f_{\rm A}$ is the AGN fraction among central galaxies at $M_{\rm DMH}\ga M_{\rm min}$ and $\Theta(x)$ is a step function that has the value of 0 at $x<0$ and 1 at $x\geq 0$ respectively. We do not introduce the smoothing of the step function through the parameter $\sigma_{\log\,M}$ for the AGN HOD because this parameter is found to be poorly constrained and pegged at zero when it is made a free parameter. Again, we fix $M_0$ to zero.
The HOD has implications for the AGN population that are satellites.  The satellite fraction $f_{\rm sat}$ is often calculated in literature from the AGN HOD analysis. The satellite fraction can be calculated by:

\begin{equation}
    f_{\rm sat}=
    \frac{\int \langle N_{\rm A,s}\rangle ( \textrm{d}\phi/ \textrm{d}\log M_{\rm DMH})\textrm{d} \log M_{\rm DMH}}{\int \left(\langle N_{\rm A,c}\rangle+\langle N_{\rm A,s}\rangle\right)( \textrm{d}\phi/ \textrm{d}\log M_{\rm DMH}) \textrm{d}\log M_{\rm DMH}},
    \label{eq:fsat}
\end{equation}

where $ \textrm{d}\phi/ \textrm{d}\log M_{\rm DMH}$ is the halo mass function.
We note that this satellite fraction depends on the form of the HOD model and should not be interpreted literally. Because of the rapidly dropping nature of the DMH mass function, $f_{\rm sat}$ is dominantly contributed by the satellite population at the low DMH mass end near $M_{\rm min}$. We calculate the satellite fraction as a rough indicator of the importance of satellites among the AGN population in $M_{\rm DMH}\sim M_{\rm min}$ halos and its comparison between satellite populations between the $z\sim 1$ and $z\sim 2$ samples.

\subsection{The $\chi^2$ minimization}

To find the galaxy and AGN HODs from the galaxy ACF and galaxy-AGN CCF, we minimize the joint $\chi^2$ to the ACF and CCF. Unlike the case of, e.g., \cite{2011ApJ...726...83M}, the errors on the galaxy HOD parameters from the ACF are not negligible. Thus we choose not to fix the galaxy HOD parameters to search for confidence ranges of the AGN HOD parameters, but we fit the galaxy ACF and galaxy-AGN CCF simultaneously by minimizing 

\begin{eqnarray}
\chi^2 = & \sum_{i,j} [(w_{{\rm p}i}(r_{{\rm p},i})-w_{{\rm p, mdl}i}(r_{{\rm p},i})] M_{ij}^{-1}\nonumber\\
&[w_{{\rm p}j}(r_{{\rm p},j})-w_{{\rm p, mdl}j}(r_{{\rm p},j})].
\label{eq:acfccfchi2}
\end{eqnarray}
See Sect. \ref{sec:error_cov} for the indices $i$ and $j$. The subscript ${\rm mdl}$ means the model value evaluated at the center of the bin. As explained in Appendix \ref{sec:app_covariance}, for our final results, we use all the diagonal elements of $M_{ij}$ and its off-diagonal elements within the CCF only, while neglecting other off-diagonal elements.
Note that the ACF model depends on the galaxy HOD parameters only while the CCF model depends on both the galaxy and AGN HOD parameters.

The $\chi^2$ minimizations and determination of the HOD parameters the confidence ranges are made with an MCMC method using the MCMC-F90 library by Marko Laine\footnote{\href{http://helios.fmi.fi/~lainema/mcmcf90/}{http://helios.fmi.fi/~lainema/mcmcf90/}}
modified by us and linked to our HOD model calculation software.

\begin{figure*}[!t]
\begin{center}
\includegraphics[bb= 30 0 870 520,clip,width=16.5cm]{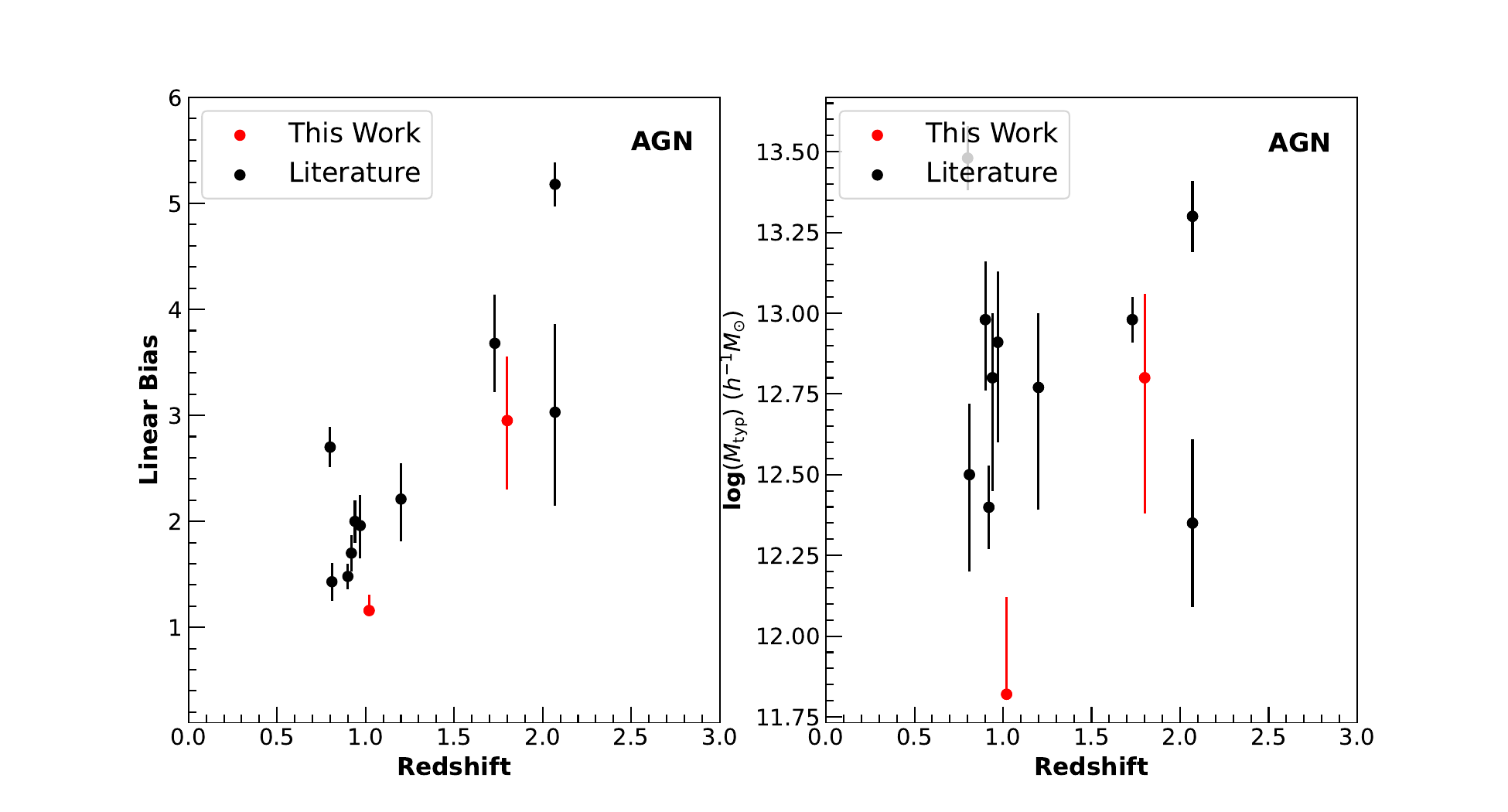}
\caption{Left: Redshift evolution of X-ray-selected AGN bias. Red and black points represent our results and results in literature ordered according to redshift (\citealt{2011ApJ...736...99A,2016MNRAS.457.4195M,2009ApJ...701.1484C,2006ApJ...645...68Y,2009A&A...494...33G,2013MNRAS.430..661M,2019A&A...629A..14V,2018A&A...620A..17P,2011ApJ...736...99A,2006ApJ...645...68Y}). Error bars of our results are at the $68\% $ confidence level (between $16\% $ and $84\% $ percentiles from the corresponding MCMC chains) for one interesting parameter. Right: Redshift evolution of log$M_{\rm typ}$ for X-ray-selected AGNs. Red and black points represent our results and results in the literature ordered according to redshift (\citealt{2011ApJ...736...99A,2016MNRAS.457.4195M,2009ApJ...701.1484C,2006ApJ...645...68Y,2009A&A...494...33G,2013MNRAS.430..661M,2019A&A...629A..14V,2018A&A...620A..17P,2011ApJ...736...99A,2006ApJ...645...68Y}). Error bars of our results are at the $68\% $ confidence level (between $16\% $ and $84\% $ percentiles from the corresponding MCMC chains) for one interesting parameter.\label{fig:agn_bias_z}}  
\end{center}
\end{figure*}

\begin{figure*}[!t]
\begin{center}
\includegraphics[bb= 40 0 820 360,clip,width=18cm]{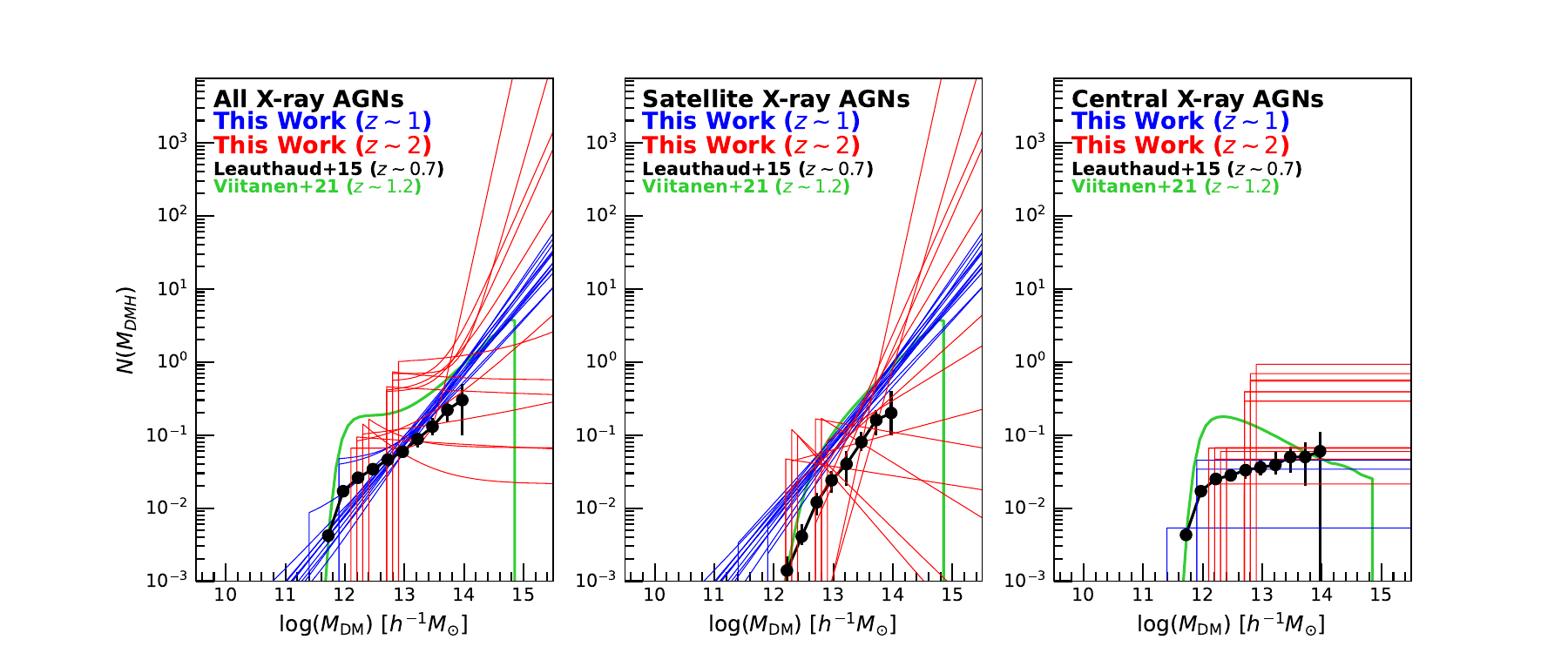}
\caption{Comparison between the HOD of X-ray-selected AGNs at $z\sim1$ and $z\sim2$. In each panel, the blue and red solid lines show the 68\% confidence ranges of central ($\langle N_{\rm A,c} \rangle$), satellite  ($\langle N_{\rm A,s} \rangle$), and the central+satellite HODs. Blue and red lines show the central+satellite HODs at $z\sim1$ and $z\sim2$, respectively. Black circles show the HOD of X-ray-selected AGNs at $z\sim0.7$, which is derived by \cite{2015MNRAS.446.1874L}. Green lines show the HOD of mock AGNs at $z\sim1.2$, which is derived by \cite{2021MNRAS.507.6148V}. \label{fig:hod_z1_z2}} 

\end{center}
\end{figure*}
\section{Results}

Using Eq. \ref{eq:wprpint} in Sect. \ref{sec:acf}, we calculate the ACF of galaxies at $0.6<z<1.4$ and $1.4<z<2.6$, respectively. 
Also, using Eq. 10 in Sect. \ref{sec:ccf}, we calculate the CCF of galaxies and AGNs at $0.6<z<1.4$ and $1.4<z<2.6$, respectively.
The jack-knife resampling technique calculates the errors of ACF and CCF. Figure \ref{fig:wprp_all} shows our ACF/CCF measurements with the best-fit models and fit residuals. The measurements of the $r_{\rm p}>10$ bin are plotted for reference, but not included in the fitting process, because this bin corresponds to angular scales comparable to or larger than those of the Jacknife zones. Due to the large error and low measured value of the smallest angular bin of the $1.4<z<2.6$ CCF, this bin is not included in the fit either. The corresponding angular scale of this bin is comparable to the {\it Chandra} PSF at larger off-axis angles and the blending of multiple sources is a likely cause of its low value.

Then we obtain constraints on the AGN HOD parameters, such as $M_{\rm min}$ and  $\alpha_{\rm s}$. Using these two parameters, we can calculate the mean DMH mass occupied by the X-ray AGN sample:

\begin{equation}
\langle M_{\rm DMH}\rangle=\frac{\int M_{\rm DMH} \langle N_{\rm A}\rangle (M_{\rm DMH}) \phi(M_{\rm DMH}) \textrm{d}M_{\rm DMH}}
                      {\int \langle N_{\rm A}\rangle (M_{\rm DMH}) \phi(M_{\rm DMH}) \textrm{d}M_{\rm DMH}},
\label{eq:meanmass}
\end{equation} 

and also the effective linear bias parameters of galaxies and AGNs is calculated respectively as follows:

\begin{eqnarray}
b_{\rm G}&=&\frac{\int b_{\rm DMH}(M_{\rm DMH})\langle N_{\rm G}\rangle (M_{\rm DMH})\phi(M_{\rm DMH}) \textrm{d}M_{\rm DMH}}
 {\int \langle N_{\rm G}\rangle(M_{\rm DMH})\phi(M_{\rm DMH}) \textrm{d}M_{\rm DMH}}\\
b_{\rm A}&=&\frac{\int b_{\rm DMH}(M_{\rm DMH})\langle N_{\rm A}\rangle (M_{\rm DMH})\phi(M_{\rm DMH}) \textrm{d}M_{\rm DMH}}
        {\int \langle N_{\rm A}\rangle (M_{\rm DMH})\phi(M_{\rm DMH}) \textrm{d}M_{\rm DMH}}.
\label{eq:bias_A}
\end{eqnarray}

The confidence contours of the HOD model parameters for galaxies at $0.6<z<1.4$ and $1.4<z<2.6$ are shown in off-diagonal panels of Figure \ref{fig:gal_hod_contours} in two-dimensional projected parameter spaces. The 68\% and 95\% confidence contours are drawn along the equal $\chi^2$ values.  For each panel in each two-parameter grid bin, we find many MCMC points, which have different values of other parameters and accordingly different $\chi^2$ values.  We pick the smallest $\chi^2$ value to represent this two-dimensional parameter bin and use these $\chi^2$ values for contouring. The two $\chi^2$ contour levels are determined such that 68\% and 95\% of all the MCMC chain points fall below these $\chi^2$ levels respectively. We also plot the histogram of MCMC points in single parameters in diagonal panels. In the same way, the confidence contours and histograms of HOD models are shown for the $0.6<z<1.4$ and $1.4<z<2.6$ X-ray-selected AGN samples in Figure \ref{fig:agn_hod_contours}. 
Table \ref{tab:hod_params} summarizes the estimated HOD parameters, the derived linear bias parameter and typical DMH mass ($M_{\rm typ}$). Here $M_{\rm typ}$ is defined by $b_{\rm DMH}(M_{\rm typ})=b_{\rm lin}$, where $b_{\rm lin}$ is the observed linear bias parameter for the sample. In Table \ref{tab:hod_params}, the best-fit value of each parameter is from the minimum $\chi^2$ model and the 68\% confidence range is shown in the parentheses, which is derived from the marginalized distributions of the MCMC chain. In rare cases, the probability distribution is skewed and the nominal value can be slightly outside of the 68\% confidence range. 

Figure \ref{fig:agn_hod_plots} (upper panels) shows random selections of acceptable AGN HOD models for the central, satellite components, and the total, where we consider models with the smallest 68\% $\chi^2$ in the MCMC chain acceptable. In figure labels and text, we express various HODs as "$N(M_{\rm DMH})$" generically. The $N(M_{\rm DMH})$ values have been normalized to have the effective space density of the sample, which is the number of the AGNs divided by the total comoving volume in the field over the redshift range (see Table \ref{tab:xraysamp}). 
About half of the initially accepted $z\sim 2$ AGN central HOD curves showed $N_{\rm A,c}>1$ (or $f_{\rm A}>1$) upon the normalization, which is unphysical. These have been excluded from the chain, and the 68\% confidence ranges are recalculated accordingly. Table \ref{tab:hod_params} (bottom row), Figure \ref{fig:agn_hod_contours} (bottom), and Figure \ref{fig:agn_hod_plots}(right panels) show the results of the chain after the exclusion. In the lower panels, we also show $N(M_{\rm DMH})$ multiplied by the halo mass function. For the $z\sim 1$ sample, the fraction of initially accepted HOD chains points that give $N_{\rm A,c}>1$ is less than 1\% and therefore neglected.

\section{Discussion}
\subsection{The bias and dark matter halo mass}

 To investigate whether the calculated galaxy bias aligns with previous measurements, we compared the galaxy bias and $\log M_{\rm typ}$ of galaxies at approximately $z = 1$ and $z = 2$ with the literature (\citealt{2011ApJ...728...46W,2018ApJ...853...69C}). The results of this comparison are illustrated in Figure \ref{fig:gal_bias_mstar}. At around $z = 1$, we found a galaxy bias of $1.50\ (1.33; 1.52)$. At $z \sim 2$, our analysis yielded a galaxy bias of $2.03\ (1.92;2.06)$.

Similar findings were reported by \cite{2011ApJ...728...46W} based on a galaxy sample from the NOAO Extremely Wide-Field Infrared Imager Medium Band Survey. Their analysis resulted in a galaxy bias of $2.08^{+0.04}_{-0.03}$ at $z \approx 1$. Furthermore,  \cite{2018ApJ...853...69C} derived a galaxy bias of $2.07 \pm 0.04$ at $z \approx 2$. Consequently, our study confirms a general consistency between the galaxy bias at $z \sim 1 - 2$ with previous measurements.


Figure \ref{fig:agn_bias_z} shows the evolution of the AGN bias parameter and the associated $M_{\rm typ}$ from this work and from the literature. From this work, the AGN linear bias $b_{\rm A}$
and $M_{\rm typ}$ are significantly larger at $z\sim 2$ than at $z\sim 1$. Since the COSMOS Legacy sample is essentially flux-limited, the $z\sim 2$ sample contains more luminous AGNs than the $z\sim 1$ sample. To trace the redshift evolution of halo masses AGNs on the same ground, we need to combine the results from different surveys.
A caveat is that each sample from the literature cover different X-ray-selected AGN selection criteria in terms of energy bands, luminosities, and the redshift range.

In order to assess the consistency of the calculated AGN bias with previous measurements, we conducted a comparison of AGN bias and the logarithm of $M_{\rm typ}$ for X-ray-selected AGNs at approximately $z\sim1$ and $z\sim2$ with reference to a study in the literature (\citealt{2009ApJ...701.1484C}). At around $z\sim1$, \textbf{we determined an inferred AGN bias of $1.16 \ (1.16;1.31)$, which corresponds to a DMH mass of log($M_{\rm typ}/M_{\odot}$) $=11.82\ (11.82; 12.12)$.}
In comparison, the AGN bias and log($M_{\rm typ}$) values for X-ray-selected AGNs at $z\sim1$, as reported in \cite{2009ApJ...701.1484C}, stand at $1.48\pm0.12$ and $12.98\ (12.76; 13.16)$, respectively. For X-ray-selected AGNs at $z\sim2$, our calculations yielded an AGN bias of $2.95\ (2.30; 3.55)$ and a log($M_{\rm typ}/M_{\odot}$) of $12.80\ (12.38; 13.06)$. In contrast, \cite{2011ApJ...736...99A} investigated AGN bias at $z\sim2$ by Equation (16) in \cite{2011ApJ...736...99A} and they estimated that AGN bias at $z\sim2$ is $5.18\pm0.21$, which correspond to log($M_{\rm typ}/M_{\odot}$) of $13.30\pm0.11$.
We note that the typical DMH mass derived from the large-scale bias may not reflect the true distribution of AGN host halo masses. This is especially the case if the underlying AGN host halo mass distribution spans a range of halo masses in different large-scale environments. For example, \citet{2015MNRAS.446.1874L} found a skewed distribution of AGN host DMH masses in COSMOS, where the typical DMH masses reported earlier resided between their median (lower) and mean (higher) values. Also, using DMHs from large N-body simulations and empirically motivated AGN
models \citet{2021MNRAS.502.5962A} found that the true bias calculated directly from the DMHs is systematically lower than the observationally measured one.  


\begin{figure*}[!t]
\begin{center}
\includegraphics[bb= 0 0 720 620,clip,width=17cm]{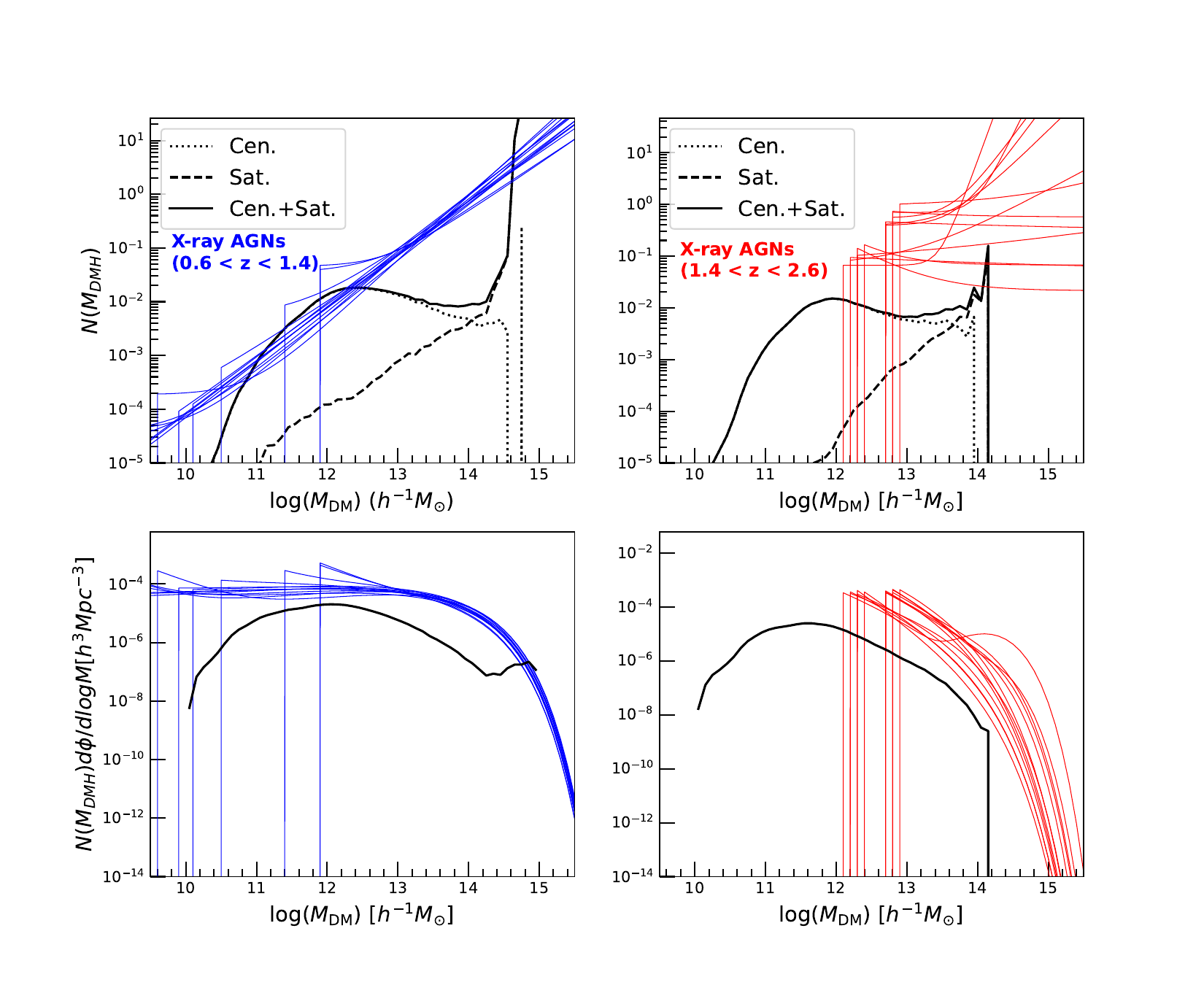}
\caption{Comparison between the HOD of our results and theoretical predictions from the Uchuu-nu2GC catalogs.  In each panel, the dotted, dashed, and solid lines show the HODs of central ($\langle N_{\rm A,c} \rangle$), satellite  ($\langle N_{\rm A,s} \rangle$) and the central+satellite HODs. Blue and red lines show the central+satellite HODs at $z\sim1$ and $z\sim2$, respectively.
\label{fig:hod_vs_semi}}
\end{center}
\end{figure*}




                                

\begin{figure*}[!ht]
\begin{center}
\includegraphics[bb= 40 0 1120 400,clip,width=20cm]{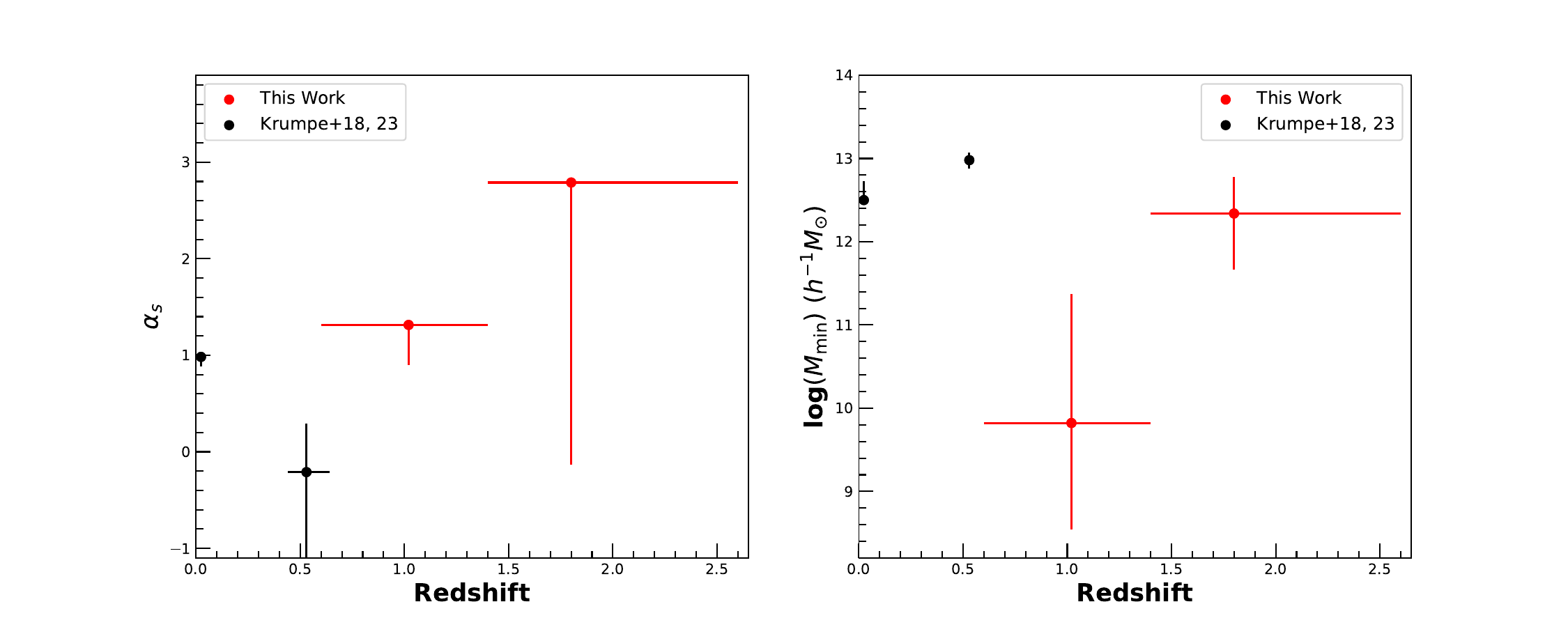}
\caption{Redshift evolution of $\alpha_{s}$ (satellite slope) and log($M_{\rm min}/M_{\odot}$). Red and black points represent our results and the results in the literature (\citealt{2018MNRAS.474.1773K,krumpe23}).  Error bars of our results are at the $68\% $ confidence level (between $16\% $ and $84\% $ percentiles from the corresponding MCMC chains) for one interesting parameter.
\label{fig:alpha_vs_z}}
\end{center}
\end{figure*}

\subsection{The HOD of X-ray-selected AGNs at $z\sim1$ and $z\sim2$}
To investigate the redshift-dependent behavior of the HODs of X-ray-selected AGNs in detail, we present the HODs of X-ray-selected AGNs at $z\sim1$ and $z\sim 2$ in Figure \ref{fig:hod_z1_z2}. The blue lines represent the HODs of X-ray-selected AGNs at $z\sim1$. Similarly, the red lines show the HODs of X-ray-selected AGNs at $z\sim2$.

Under this form of HOD model (Eqs. ~\ref{eq:agnhodcen} \& ~\ref{eq:agnhodsat}), there are two peaks in the distribution of $f_{\rm sat}$ for $z\sim 1$ AGNs, where all or mostly central models are highly unlikely, with only 1.4\% of the points in the MCMC chain give $f_{\rm sat}<0.05$. At $z\sim 2$, the HOD analysis suggests a small fraction of satellites, with all/mostly central models are highly unlikely, with only 1.5\% of the MCMC chain points give $f_{\rm sat}>0.95$. (See the histogram of the lower right corner in each of Figure \ref{fig:agn_hod_contours}.).
{At these redshifts, the halo mass function drops rapidly at group/cluster-sized halos, thus there are practically no $z\sim 2$ rich groups/clusters. Therefore, we focus on the discussion on $\log M_{\rm DMH}<13$.}

\cite{2015MNRAS.446.1874L} measured the X-ray-selected AGN HOD at $z\sim0.7$ in COSMOS using weak gravitational lensing. \cite{2021MNRAS.507.6148V} also derived HODs of mock AGNs at $z\sim1.2$ with different $Q$ values. They define $Q=U_{\rm sat}/U_{\rm cen}$, where $U_{\rm sat}$ and $U_{\rm cen}$ are the duty cycle for the satellite and central BHs, respectively. Since the HOD of mock AGNs at $z\sim1.2$ with $Q=1$ agrees with that of other studies (e.g., \citealt{2021MNRAS.502.5962A}), we show it as a representative example of HODs derived by \cite{2021MNRAS.507.6148V} in Figure 10. The HOD of our results at $z\sim1$ is generally consistent with that of \cite{2015MNRAS.446.1874L} and \cite{2021MNRAS.507.6148V}.

We perform a comparison between our findings and the theoretical results presented by \cite{2020MNRAS.497....1O}. \cite{2020MNRAS.497....1O} have predicted the X-ray-selected AGN bias using semi-analytic modeling of AGNs and have verified that their results are broadly consistent with observational findings. Our results are also in good agreement with the predictions made by \cite{2020MNRAS.497....1O}.
We compare our HOD results with theoretical predictions. To facilitate this comparison, we have constructed HODs for theoretical predictions at $z\sim2$. Since the $z\sim 2$ HOD itself is not presented by \cite{2020MNRAS.497....1O}, we generated the HODs at $z\sim1$ and $z\sim2$ using the Ucchu + nu2GC galaxies and AGN catalog (\citealt{2023MNRAS.525.3879O}). This catalog is generated by utilizing the Uchuu simulation that is a 2.1 trillion N-body simulation that uses Planck cosmology (\citealt{2021MNRAS.506.4210I}). The comparison between our observational results and the theoretical predictions is shown in Figure \ref{fig:hod_vs_semi}.
At $z \sim 1$, our observational data aligns remarkably well with the theoretical predictions, falling within the margins of error.

Next, we show the redshift evolution of $\alpha_{\rm s}$ in the left side of Figure \ref{fig:alpha_vs_z}. \cite{2011ApJ...726...83M, 2018MNRAS.474.1773K,krumpe23} conducted HOD modeling of X-ray-selected AGNs at $z<1$ and reported $0\la\alpha_{\rm s}\la1$. In contrast, our analysis estimates $\alpha_{\rm s}$ at $z\sim1-2$, revealing a trend of increasing $\alpha_{s}$ with redshift, with $\alpha_{s}$ at $z\sim1-2$ being greater than 0. As \cite{krumpe23} mentioned, $\alpha_{\rm s} > 0$ indicates that the number of AGNs in satellite galaxies increases with $M_{\rm DMH}$. Conversely, $\alpha_{\rm s} < 0$ suggests that AGNs in satellite galaxies are more commonly found in lower-mass halos and are infrequent in high-mass halos. Therefore, the results of this study suggest that the number of AGNs in satellite galaxies gradually increases with $M_{\rm DMH}$ up to $z=2$. We also show the redshift evolution of log($M_{\rm min}/M_{\odot}$) on the right side of Figure \ref{fig:alpha_vs_z}. We find that log($M_{\rm min}/M_{\odot}$) of X-ray-selected AGNs at $z<1$ is $\sim12.5-13.0$, but that of X-ray-selected AGNs at $z\sim1-2$ is $<12.5$. 

\section{Summary}
We have calculated the ACF of galaxies and the CCF of galaxies and X-ray-selected AGNs at $0.6<z<2.6$ in the COSMOS field to estimate the HOD of X-ray-selected AGNs up to $z\sim2$.
The main results of our work are summarized below.
\begin{itemize}
\item We have performed the HOD modeling of X-ray-selected AGNs by utilizing the ACF of galaxies and CCF of galaxies and AGNs at $z\sim1-2$.
\item We have estimated AGN bias values of $b=1.16\ (1.16;1.31)$ and $b=2.95\ (2.30;3.55)$, respectively. These values correspond to typical host DMH masses of log$(M_{\rm typ}/M_{\odot})=11.82\ (11.82;12.12)$ and log$(M_{\rm typ}/M_{\odot})=12.80\ (12.38;13.06)$, respectively.
\item We have found a significant satellite AGN population at $z\sim 1$ all over the DMH mass range occupied by AGNs.  While $z\sim 2$ AGNs in our sample are associated with higher mass DMHs, the satellite fraction is smaller.
\item The HOD analysis suggests a marginal tendency of increasing $\alpha_{\rm s}$ with redshift, but larger samples are needed to confirm this with sufficient statistical significance.
\item We have found that the best-fit values of $\alpha_{\rm s}$ in both redshift bins are greater than 0, suggesting tendencies of increasing satellite AGN number with $M_{\rm DMH}$.
\end{itemize}

\begin{acknowledgments}
We thank Mara Salvato for providing us with the PDz's for the X-ray source counterparts of the {\sl Chandra} COSMOS Legacy Survey.
We thank Instituto de Astrofisica de Andalucia (IAA-CSIC), Centro de Supercomputacion de Galicia (CESGA) and the Spanish academic and research network (RedIRIS) in Spain for hosting Uchuu DR1, DR2 and DR3 in the Skies \& Universes site for cosmological simulations. The Uchuu simulations were carried out on Aterui II supercomputer at Center for Computational Astrophysics, CfCA, of National Astronomical Observatory of Japan, and the K computer at the RIKEN Advanced Institute for Computational Science. The Uchuu Data Releases efforts have made use of the skun@IAA\_RedIRIS and skun6@IAA computer facilities managed by the IAA-CSIC in Spain (MICINN EU-Feder grant EQC2018-004366-P). Data analysis was in part carried out on the Multi-wavelength Data Analysis System operated by the Astronomy Data Center (ADC), National Astronomical Observatory of Japan.

This work is supported by UNAM-DGAPA PAPIIT IN111319, IN114423 and CONACyT Grant Cient\'ifica B\'asica 252531 (TM). This work is also supported by JSPS KAKENHI Grant numbers 23K03465 (HI), 22H01266 (YT), 21H05449, and 23K03460 (TO). TM thanks JSPS for financial support under the "Invitational Fellowship for Research in Japan (L24523)" program and Yoshihiro Ueda and Kyoto University for their hospitality during his sabbatical stay.
\end{acknowledgments}

\appendix
\section{Covariance Matrix}
\label{sec:app_covariance}
The errors in different bins of the correlation functions are mutually correlated and the $\chi^2$ in the presence of the correlated errors can be treated with the covariance matrix  (Eq. \ref{eq:cov}) constructed from Jackknife resamplings with Eq. \ref{eq:acfccfchi2}. For accurate measurements of the covariance matrix, a large number of Jackknife zones that are statistically independent from one another would be ideal.  However, the number of the Jackknife zones is limited by the size of each zone, which should be larger than the largest $r_{\rm p}$ scale to be measured so that each zone is as statistically independent as possible. Strictly speaking, when there are pairs across different zones, the statistical independence is not met to some degree. This is a fundamental limitation of using this method.

\begin{figure*}[!ht]
\begin{center}
\includegraphics[height=0.45\columnwidth]{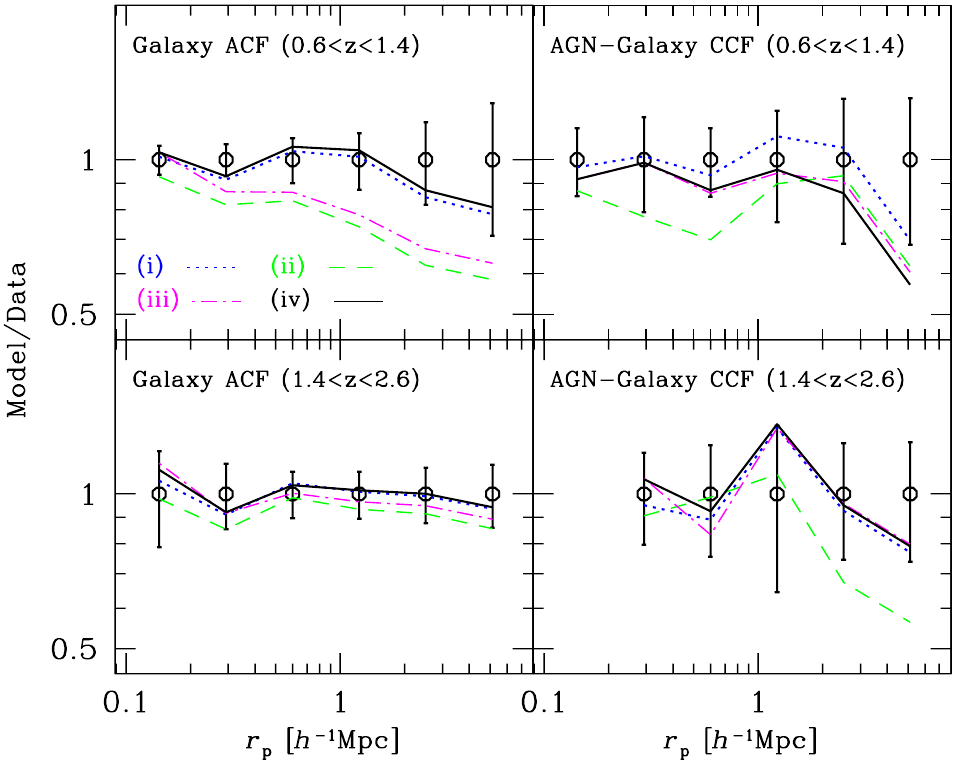}
\includegraphics[height=0.45\columnwidth]{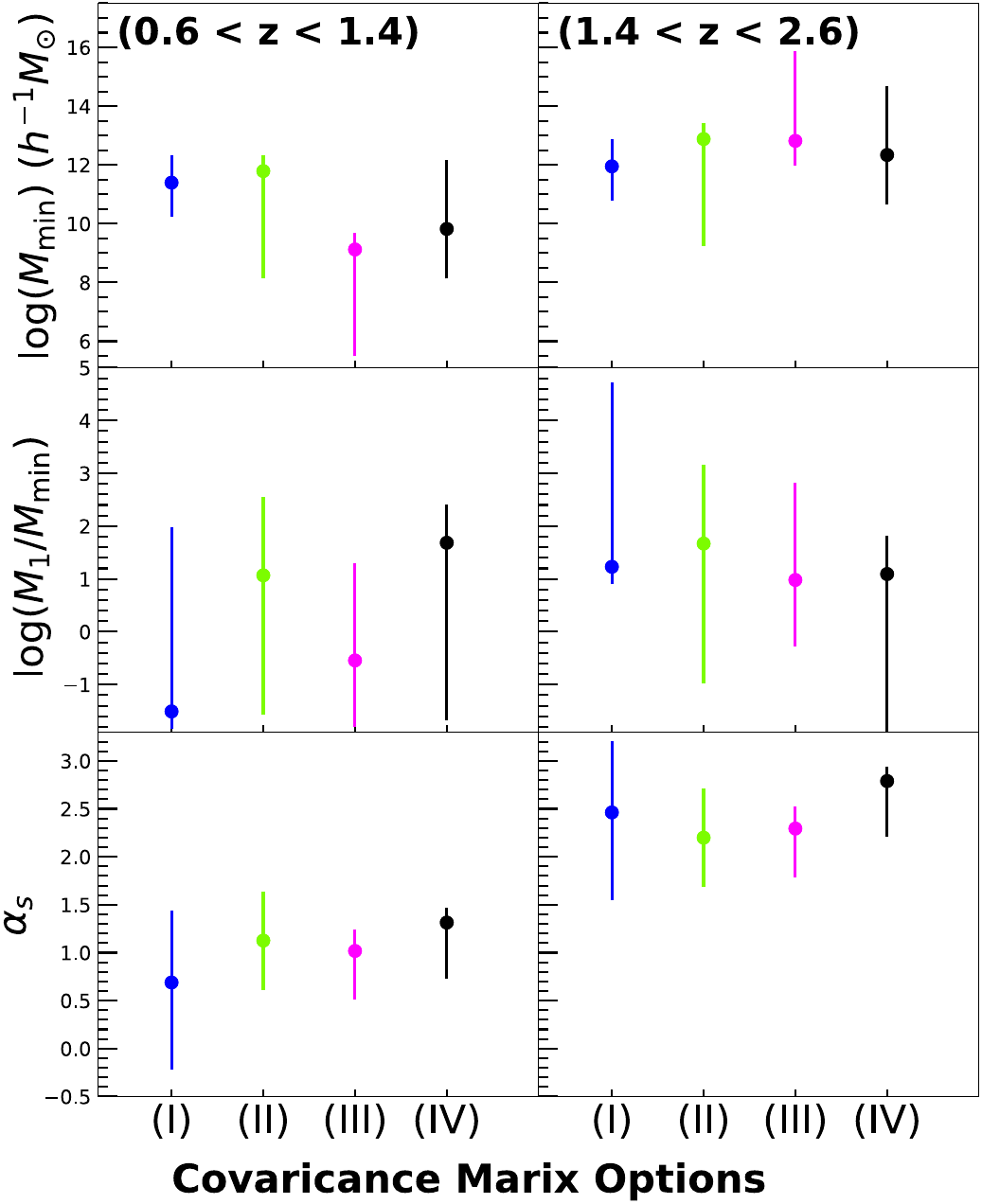}
\end{center}
\caption{{\it Left:} Ratios of the $w_{\rm p}(r_{\rm p})$ models giving minimum $\chi^2$ with different covariance matrix options to the measured values. The circles with 1$\sigma$ error bars represent measurements used for the fits. The blue-dotted, green-dashed, magenta dot-dashed and black-solid lines are the models divided by the measured values corresponding to the covariance matrix options (i)-(iv) (See text). {\it Right:} Minimum $\chi^2$ parameter values for the AGN HODs and assocaited 68\% errors for the covariance-matrix options (i)-(iv).
\label{fig:comp_covs}}
\end{figure*}

In the case of a small field such as ours, we can divide the field to only a small number of Jackknife zones (16 in our case), and some off-diagonal regions of the covariance matrix become noise-dominated \citep[e.g.][]{HerreoAlonso23}. This can cause the model with the minimum correlated $\chi^2$ to significantly deviate from the data. To avoid this problem, we clean the covariance matrix by ignoring some off-diagonal parts. We have tried the following covariance matrices: (i) uncorrelated $\chi^2$ (all off-diagonal elements are zeros), (ii) full covariance matrix, keeping all off-diagonal elements including those between different $r_{\rm p}$ bins within each of the galaxy ACF and galaxy-AGN CCF as well as elements across them, (iii) the off-diagonal elements across the ACF and CCF are set zeros while keeping those within each, and (iv) only off-diagonal elements within the CCF are kept.  
Figure \ref{fig:comp_covs} {\it (left)} shows the ratios of actual measurements with the models that give minimum $\chi^2$ using covariance matrices (i)-(iv). Naturally, using uncorrelated $\chi^2$ (i) gives the best apparent fits to the data. However, under the existence of the correlations among the data errors, the parameter errors/confidence ranges obtained using (i) may well be misestimated. The correct method is to use the full covariance matrix (ii) if it is accurate enough. However, we observe that the best-fit model using this option significantly underestimates the 0.6$<z<$1.4 ACF at $r_{\rm p}\ga 1 h^{-1} [{\rm Mpc}]$ and 1.4$<z<$2.6 CCF at $r_{\rm p}\ga 2 h^{-1} [{\rm Mpc}]$.
Similar deviations remain for the 0.6$<z<$1.4 ACF even if the off-diagonal elements across the ACF/CCF are set to zero. Yet another option is to keep off-diagonal elements within the CCF, while setting the others zeros (option iv). As seen in Figure \ref{fig:comp_covs} {\it (left)}, the best-fit models with this option are almost identical to those with option (i) for both redshift ranges. Figure \ref{fig:comp_covs} {\it (right)} compares the AGN HOD parameters that give minimum $\chi^2$ and confidence ranges estimated by the marginal distributions of MCMC chain point for options (1)-(iv). This panel shows that the option (i) (uncorrelated $\chi^2$) gives distinctively different errors for some parameters, while those from other options give approximately same. Option (iv) (only off-diagonal elements within the CCF are kept) does not show significant residuals of the fit at the same time, gives errors similar to 
those from option (ii) (full covariance matrix). Thus, in this work, we choose to use the option (iv) for further analysis.

\bibliography{adssample}

\end{document}